\documentclass[letterpaper,twocolumn,10pt]{article}
\usepackage{usenix}
\usepackage[available, functional]{usenixbadges}
\usepackage{pifont}
\usepackage{placeins}
\usepackage{pgfplotstable}
\usepackage{paralist}
\usepackage{amsmath}
\usepackage{multirow}
\usepackage{shortcuts}
\usepackage{epsfig,endnotes}
\usepackage{tikz}
\usepackage{pgfplots}
\usepackage{listings}
\usepackage[plainruled, linesnumbered, noline]{algorithm2e}
\usepackage{xspace}
\usepackage{amsfonts}
\usepackage{url}
\usepackage{tikz}
\usepackage{array}
\usepackage{booktabs,adjustbox}
\usepackage{etex}
\usepackage{color}
\usepackage{pstricks}
\usetikzlibrary{calc,positioning,shapes.geometric,shapes.symbols,shadows,arrows}
\usepackage{wasysym}
\usepackage{shortcuts}
\usepackage{paralist}
\renewcommand{\paragraph}[1]{\vspace{0.05in}\noindent{\bf{#1}.}}
\usepackage[justification=justified]{caption}
\usepackage{pgfplots}
\usetikzlibrary{arrows}
\usetikzlibrary{patterns}
\usepackage{times}
\usepackage{listings}
\usepackage{multicol}
\usepackage{lipsum}
\usepackage{adjustbox}
\usepackage[font=bf,skip=\baselineskip]{caption}
\usepackage{tabularx}
\usepackage{multirow}
\usepackage{arydshln}

\author{
{\rm Chuyang Chen}\\
\it The Ohio State University \\
\it chen.13875@osu.edu
\and
{\rm Brendan Dolan-Gavitt}\\
\it New York University \\
\it brendandg@nyu.edu
\and
{\rm Zhiqiang Lin}\\
\it The Ohio State University\\
\it zlin@cse.ohio-state.edu
} 
\date{}
\usepackage{titling}
\usepackage{hyperref}
\usepackage{xcolor}
\hypersetup{
    colorlinks,
    linkcolor={red!50!black},
    citecolor={blue!50!black},
    urlcolor={blue!80!black}
}
\usepackage{graphicx}
\usepackage{paralist}
\usepackage{wrapfig}
\usepackage[frozencache,cachedir=.]{minted}
\setminted{fontsize=\footnotesize}

\usepackage[english]{babel}
\usepackage{color,soul}
\usepackage[numbers,sort&compress]{natbib}
\usepackage{subcaption}
\usepackage[most]{tcolorbox}
\usepackage{listings}

\usepackage{tikz}
\usepackage{rotating}
\usepackage{amsthm}
\usepackage{appendix}
\usepackage{slantsc}
\rmfamily \DeclareFontShape{OT1}{ptm}{m}{scit}{<->ssub * lmr/m/scsl}{} \DeclareFontShape{OT1}{ptm}{bx}{scit}{<->ssub * lmr/bx/scsl}{}

\usepackage{hyperref}
\usepackage{float}

\newfloat{lstfloat}{htbp}{lop}
\floatname{lstfloat}{Listing}

\hypersetup{
    colorlinks = true,
    linkcolor = blue,
    anchorcolor = blue,
    citecolor = blue,
    filecolor = blue,
    urlcolor = blue
}

\DeclareMathOperator*{\argmax}{arg\,max}

\usepackage[all]{nowidow}

 \usepackage{caption}
\lstdefinestyle{tiny}{
    basicstyle=\tiny,
}

\theoremstyle{definition}
\newtheorem{definition}{Definition}

\newcommand{\markone}{\mbox{\color[HTML]{666666}{\normalsize\ding{182}\xspace}}}
\newcommand{\marktwo}{\mbox{\color[HTML]{FF7D00}{\normalsize\ding{183}\xspace}}}
\newcommand{\markthree}{\mbox{\color[HTML]{1F8FFF}{\normalsize\ding{184}\xspace}}}
\newcommand{\markfour}{\mbox{\color[HTML]{4F7D29}{\normalsize\ding{185}\xspace}}}
\newcommand{\markfive}{\mbox{\color[HTML]{8C008C}{\normalsize\ding{186}\xspace}}}
\newcommand{\marksix}{\mbox{\color[HTML]{001D77}{\normalsize\ding{187}\xspace}}}
\newcommand{\markseven}{\mbox{\color[HTML]{B21C1C}{\normalsize\ding{188}\xspace}}}
\newcommand{\markyes}{\ding{51}}
\newcommand{\markno}{}

\definecolor{mygray}{HTML}{CCCCCC}

\soulregister\cite7
\soulregister\ref7
\soulregister\sysnamehl0
\soulregister\codespanalt1
\soulregister\autoref7
\soulregister\pageref7

\newcommand{\markch}[1]{#1}

\usepackage[all]{nowidow}
\newcommand{\codespan}[1]{
    \fbox{\scriptsize{\texttt{\vphantom{(*}#1}}}
}

\newcommand{\codespanalt}[1]{
    \fbox{\scriptsize{\texttt{\vphantom{T}#1}}}
}

\SetAlCapSkip{3pt plus 1pt minus 2pt}
\begin{document}
\title{{\scshape ELFuzz}: Efficient Input Generation via LLM-driven Synthesis Over\\ Fuzzer Space}

\maketitle

\begin{abstract}
    Generation-based fuzzing produces appropriate testing cases according to specifications of input grammars and semantic constraints to test systems and software. However, these specifications require significant manual effort to construct. This paper proposes a new approach, \caps{ELFuzz} (\underline{E}volution Through \underline{L}arge Language Models for \underline{Fuzz}ing), that automatically synthesizes generation-based fuzzers tailored to a system under test (SUT) via LLM-driven synthesis over fuzzer space. At a high level, it starts with minimal seed fuzzers and propels the synthesis by fully automated LLM-driven evolution with coverage guidance. Compared to previous techniques, \caps{ELFuzz} can 1) seamlessly scale to SUTs of real-world sizes (up to 1,791,104 lines of code in our evaluation) and 2) synthesize efficient fuzzers that catch interesting grammatical structures and semantic constraints in a human-understandable way. Our evaluation shows that \caps{ELFuzz} achieves up to 434.8\% more coverage over the second best and triggers up to 216.7\% more artificially injected bugs, compared to the state-of-the-art. We also used \caps{ELFuzz} to conduct a real-world fuzzing campaign on the newest version of \texttt{cvc5} for 14 days, and encouragingly, it found five 0-day bugs (three are exploitable). Moreover, our ablation study shows that fuzzer space, the key component of \caps{ELFuzz}, contributes the most (up to 62.5\%) to the effectiveness of \caps{ELFuzz}. Further analysis of the fuzzers synthesized by \caps{ELFuzz} confirms that they correctly express the grammatical structures and semantic constraints of a SUT. The results present a promising potential of \caps{ELFuzz} for more automated, efficient, and extensible input generation for fuzzing.
\end{abstract}

\section{Introduction}
\label{sec:intro}
Over the past decade, fuzzing has proven to be a highly effective technique for finding vulnerabilities in critical software and systems such as the Linux kernel, OpenSSL, FFmpeg, and SQLite~\cite{afl, american, manes2019art}. \markch{However, fuzzing systems under test (SUTs)~\cite{system} that consume complex text formats is particularly challenging, as they impose rigorous grammatical and semantic checks on the inputs.} A specific fuzzing technique that targets these SUTs is generation-based fuzzing, where test cases are generated according to the specifications of the SUT to ensure that they pass the early-stage checks~\cite{chen2018systematic, manes2019art, songreview}. \markch{Then, the generated test cases can be fed to a mutation-based fuzzer (like AFL++~\cite{fioraldi2020afl}) as seeds to reach deeper into the codebase.} Unfortunately, constructing such specifications is extremely expensive. For example, {\sc Csmith}~\cite{yang2011finding}, a notable generation-based fuzzer targeting C compilers, consists of 80k lines of code (LoC) to meet the requirements of the C99 standard~\cite{chen2021one, xia2024fuzz4all}. Putting similar labor for every SUT in different domains would be impractical. \looseness=-1


\paragraph{The problem} Automatic synthesis of these specifications, usually expressed as grammars with semantic constraints, is a long-standing difficult problem~\cite{steinhofel2024language-based, kulkarni2021learning}. Various approaches, including data-flow and control-flow analyses, have been proposed for synthesizing grammars~\cite{gopinath2020mining, hoschele2016mining}, as well as template-based techniques for synthesizing semantic constraints~\cite{steinhofel2022input}. However, these approaches are far from being practical for real-world SUTs due to the following two reasons: \looseness=-1
\begin{packeditemize}
    \item \textbf{Scalability.} Existing techniques rely on complex program analyses to reconstruct input grammars, but these analyses often lack engineering and algorithmic scalability. \looseness=-1
    \item \textbf{Efficiency.} Even provided with a pre-existing grammar, a fuzzer has to instantiate the grammatical rules to concrete test cases, which causes significant overhead that reduces fuzzing efficiency.
\end{packeditemize}

\paragraph{Our solution} To advance the state of the art, this paper introduces \sysname (\underline{E}volution Through \underline{L}arge Language Models for \underline{Fuzz}ing), which employs a large language model (LLM) to conduct LLM-driven synthesis over fuzzer space (explained later) to synthesize generation-based fuzzers tailored to a SUT. \sysname starts with a na\"{i}ve seed fuzzer that generates purely random texts and propels the synthesis by fully automated LLM-driven evolution with fuzzer space guidance. Compared to previous techniques, \sysname can 1) seamlessly scale to SUTs of a real-world size (up to 1,791,104 LoC in our evaluation) and 2) synthesize input generators that directly embed grammars with semantic constraints, eliminating the overhead associated with grammar rule instantiation. \looseness=-1

\paragraph{LLM-driven evolution} \sysname synthesizes the fuzzers via an LLM-driven evolution loop. The loop starts from a na\"{i}ve seed fuzzer and gradually improves towards superior fuzzers. In each evolution iteration, \sysname queries the LLM to mutate the current survivors. \markch{The knowledge about the SUT learned in training enables the LLM to conduct natural, human-like modifications (e.g., reworking several lines while maintaining consistency with the context).} Later, the mutants will be assessed according to their coverage information, and the best ones will be chosen for the next iteration. By gradually improving, the challenging task of synthesizing a whole fuzzer is divided into much simpler, small steps. Therefore, \sysname can leverage the coding capability of the LLM to ``conquer'' these tractable steps.

\paragraph{Fuzzer space guidance} \markch{The evolution loop is guided by coverage information consisting of the exact range of the covered code. Not like coverage represented by a single value that coarsely ``equalizes'' fuzzers covering the same \emph{size} of code, \sysnamehl compares the candidate fuzzers (produced by mutation during the evolution) in a fine-grained way by the \emph{ranges} of code they cover.} The fuzzer space structures the candidate fuzzers accordingly into a lattice, namely, the \textit{fuzzer space}, which is formed by the partial order relation between the range of the covered code of each fuzzer. Fuzzers that cover the same code are essentially equivalent, while those that cover more code are superior; incomparable ones (possibly with the same coverage value), on the contrary, test different parts of the SUT and cannot be substituted by one another. \markch{Fuzzer space enables fine-grained comparison of the fuzzers, which is impossible if using a single coverage value, where two fuzzers incomparable in the fuzzer space can have the same coverage value and be wrongly regarded as equivalent.} Through such evaluation and comparison, the fuzzer space guides the evolution loop toward better fuzzers.

\paragraph{Encouraging results} We have implemented and evaluated \sysname. Our evaluation shows that fuzzers synthesized by \sysname achieve 418.5\% more coverage than state-of-the-art techniques and trigger up to 216.7.0\% more artificially injected bugs. In a real-world bug-finding experiment, it found five 0-day bugs of \texttt{cvc5}, three of which are exploitable vulnerabilities. Note that two of the bugs have been fixed before our responsible disclosure, and we, therefore, only reported the other three bugs to the developers. An ablation study shows that the fuzzer space model contributes the most (up to 62.5\%) to the effectiveness of \sysname. \markch{Two case studies further demonstrate that the fuzzers synthesized by \sysnamehl catch interesting grammatical structures and semantic constraints in an interpretable way and can be easily extended with other generation-based fuzzing techniques.} \looseness=-1

\paragraph{Contributions} We make the following contributions:
\begin{packeditemize}
    \item {\bf Novel notion (\S\ref{sec:overview}).} We present the notion of fuzzer space to model the task of synthesizing generation-based fuzzers. The notion of fuzzer space provides a formal way to analyze and guide the synthesis process.
  
    \item {\bf Innovative approach (\S\ref{sec:design}).} We propose LLM-driven synthesis over fuzzer space to synthesize effective and efficient generation-based fuzzers that produce initial seeds for a mutation-based fuzzer, which further mutates them to reach deeper code. The synthesis approach is based on an LLM-driven evolution loop, which decomposes the target of synthesizing a complex fuzzer into tractable, small steps and leverages an LLM to make gradual improvements guided by the fuzzer space. We have implemented our approach as \sysnamehl and made it publicly available online (download link in \hyperref[open_science]{the open science statement}). \looseness=-1

    \item {\bf Empirical evaluation (\S\ref{sec:eval}).} We have evaluated \sysname on seven widely used benchmarks. It shows that \sysname outperforms the state-of-the-art techniques. The initial seeds provided by \sysname significantly boost later mutation-based fuzzing. Future research may investigate how to extend the LLM-driven evolution to the entire fuzzing cycle to maintain corpus diversity throughout the whole process. \looseness=-1
\end{packeditemize}

\section{Background and Related Work}
\label{sec:back}
\vspace{-0.1in}

\paragraph{Grammar-based fuzzing} Fuzzing (or fuzz testing) triggers bugs and vulnerabilities in SUTs via generating numerous random test cases~\cite{fuzzingbook2024, chen2018systematic, mallissery2023demystify, manes2019art, songreview, zhao2024systematic, zhu2022fuzzing}. While mutation-based fuzzing derives new test cases from existing ones, generation-based fuzzing synthesizes them according to specifications that define the acceptable structures~\cite{godefroid2008grammar-based, sargsyan2018grammar-based, hodovan2018grammarinator, songreview, steinhofel2022input} of the inputs to a SUT. Hardcoding such specifications in a fuzzer results in a staggering workload: \textsc{Csmith}, a notable generation-based fuzzer, consists of 80k LoC to meet the requirements of the C99 standard~\cite{chen2021one, liu2023nnsmith, xia2024fuzz4all}. \looseness=-1

Grammar-based fuzzing reduces the workload by developing a general fuzzing framework that can produce test cases for an arbitrary SUT with the specification of its inputs expressed in a domain-specific language (DSL), typically using Backus-Naur form grammars. Thus, the user only needs to provide the specification, rather than writing a complete fuzzer. However, challenges exist even in these specifications themselves: \looseness=-1
\begin{packeditemize}
    \item \textbf{Obtaining the grammar.} While DSLs mitigate the coding burden, using them to express grammars still demands considerable expertise and domain knowledge.
    \item \textbf{Semantic constraints.} Some input formats require semantic constraints beyond the expressive power of a grammar, such as the ``define-before-use'' rule in C/C++. The absence of a simple and efficient way to impose such semantic constraints impairs the quality of the generated test cases.
\end{packeditemize}

Grammar synthesis techniques are proposed to resolve the first challenge~\cite{hoschele2016mining, gopinath2020mining}. However, the complex static and dynamic analyses they used limit their engineering and algorithmic scalability. \textsc{GLADE}~\cite{bastani2017synthesizing} claims to overcome these problems. Yet, a replication study~\cite{bendrissou2022synthesizing} raises doubts about the reliability of the performance reported in the paper. The challenge of semantic constraints gains less attention. \textsc{ISLa}~\cite{steinhofel2022input} introduces an SMT-based DSL to express and solve the constraints, but the DSL is of restricted expressive power. Complementing \textsc{ISLa}, \textsc{ISLearn} proposed in the same paper mines constraints in the DSL if the templates apply.

\paragraph{Fuzz driver synthesis} Fuzz drivers take test cases generated by a fuzzer and feed them into APIs of a library SUT. They can be crucial for the fuzzing effectiveness for libraries, and many works aim to synthesize fuzz drivers automatically~\cite{liu2024oss-fuzz-gen, zhang2024how, zhang2021intelligen, jeong2023utopia}. This line of research is parallel to the target of \sysname: A fuzz driver aims to feed given test cases into a SUT, while fuzzers synthesized by \sysname aim to generate inputs to explore the SUT through a given fuzz driver. Thus, the techniques used in these works cannot be adapted to our purpose.\looseness=-1

\paragraph{Fuzzing by LLM} The emergence of LLMs has opened up unprecedented opportunities and challenges for cybersecurity research~\cite{hou2023large, vaithilingam2022expectation,  motlagh2024large, xu2024large, liu2023your}. LLMs overcome the shortcomings of previous deep learning-based fuzzing techniques (which require re-training the model on massive samples for every new SUT~\cite{godefroid2017learn, amershi2019software, wang2020systematic}) with their state-of-the-art zero-shot and few-shot learning abilities. However, fuzzing directly via LLMs poses high requirements for GPU usage during the entire fuzzing process and lacks interpretability and extensibility. \looseness=-1

\begin{table}[t]
    \centering
    \scriptsize
    \begin{tabularx}{\linewidth}{l>{\raggedleft\arraybackslash}X}
        \toprule
        \textbf{Marker} & \textbf{Challenge} \\
        \midrule
        \markone & Considering input grammars  \\
        \marktwo & Considering semantic constraints \\
        \markthree & Scaling to real-world SUTs \\
        \markfour & Preventing grammar rule instantiation \\
        \markfive & Human-understandable fuzzing process \\
        \marksix & Easily generalizable to different domains \\
        \markseven & No need for domain expertise \\
        \bottomrule
    \end{tabularx}
    \caption{Challenges for effective input generation}
    \label{tab:challenge}
\end{table}

\begin{table}[t]
    \centering
    \scriptsize
    \begin{tabularx}{\linewidth}{lr*7{>{\centering\arraybackslash}X}}
        \toprule
        \textbf{Approach} & \textbf{Year} &\markone&\marktwo&\markthree&\markfour&\markfive&\marksix&\markseven \\
        \midrule
        \textsc{AUTOGRAM}~\cite{hoschele2016mining} & 2016 & \markyes & \markno & \markno & \markno & \markyes & \markyes & \markyes \\
        \textsc{Learn\&Fuzz}~\cite{godefroid2017learn} & 2017 & \markyes & \markyes & \markyes & \markyes & \markno & \markno & \markyes \\
        \textsc{GLADE}~\cite{bastani2017synthesizing} & 2017 & \markyes & \markno & \markyes & \markno & \markyes & \markyes & \markyes \\
        \textsc{Mimid}~\cite{gopinath2020mining} & 2020 & \markyes & \markno & \markno & \markno & \markyes & \markyes & \markyes \\
        \textsc{ISLearn}~\cite{steinhofel2022input} & 2022 & \markno & \markyes & \markno & \markno & \markyes & \markno & \markno \\
        \textsc{FuzzNG}~\cite{bulekov2023no} & 2023 & \markyes & \markyes & \markyes & \markyes & \markyes & \markno & \markno \\
        \textsc{TitanFuzz}~\cite{deng2023large} & 2023 & \markyes & \markyes & \markyes & \markyes & \markno & \markno & \markno \\
        \textsc{Fuzz4All}~\cite{xia2024fuzz4all} & 2024 & \markyes & \markyes & \markyes & \markyes & \markno & \markno & \markyes \\
        \textsc{MetaMut}~\cite{ou2024mutators} & 2024 & \markyes & \markyes & \markyes & \markyes & \markyes & \markno & \markno \\
        \textsc{DY Fuzzing}~\cite{ammann2024dy} & 2024 & \markyes & \markyes & \markyes & \markyes & \markyes & \markno & \markno \\
        \textsc{SyzGen++}~\cite{chen2024syzgen} & 2024 & \markyes & \markyes & \markyes & \markyes & \markyes & \markno & \markno \\
        \textsc{CovRL-Fuzz}~\cite{eom2024fuzzing} & 2024 & \markyes & \markyes & \markyes & \markyes & \markno & \markno & \markno \\
        \textsc{FuzzInMem}~\cite{liu2024fuzzinmem} & 2024 & \markyes & \markyes & \markyes & \markyes & \markno & \markno & \markno \\
        \midrule
        \sysname & & \markyes & \markyes & \markyes & \markyes & \markyes & \markyes & \markyes \\
        \bottomrule
    \end{tabularx}
    \caption{Comparison with works targeting the challenges}
    \label{tab:related_work}
\end{table}

\paragraph{Challenges} \autoref{tab:challenge} summarizes the mentioned challenges for efficient input generation, and \autoref{tab:related_work} lists recent works that target these challenges. However, none of these techniques resolves them all. Therefore, these techniques cannot fulfill automated, efficient, and interpretable input generation. \looseness=-1

\paragraph{Our insights}  \sysname overcomes all the challenges through several novel designs:
\begin{packeditemize}
    \item \markch{\textbf{High efficiency by no grammar instantiation}. Instead of grammars, \sysnamehl synthesizes fuzzers tailored for each SUT, which skip parsing and instantiating grammar rules and generate test cases directly. The related overhead is thus avoided.} \looseness=-1
    \item \textbf{Input generation by code instead of LLMs.} \sysname provides interpretable and extensible fuzzers written in Python instead of relying on an LLM to generate test cases in a black-box manner. \looseness=-1
    \item \textbf{Decomposing the synthesis via evolution.} Considering the complexity of the task, synthesizing fuzzers could be challenging for LLMs despite their good performance on coding benchmarks~\cite{zhang2024how, vikram2024can}. Thus, \sysname adopts an evolution loop to decompose the synthesis into tractable, small steps. Such a strategy reduces the need for complex prompt engineering and powerful yet resource-intensive ``big models'' with a massive number of parameters. \looseness=-1
    \item \textbf{Fuzzer space guidance.} The evolution loop is guided by a lattice structure, namely, the fuzzer space, which is derived from the coverage information consisting of the covered code range of the candidate fuzzers. The fuzzer space guidance enables fine-grained evaluation of fuzzers while maintaining scalability. \looseness=-1
\end{packeditemize}

\section{The Notion of Fuzzer Space}
\label{sec:overview}
Fuzzer space structures fuzzers into a lattice that characterizes their effectiveness (or \textit{strength}). The process of synthesizing good fuzzers is modeled as exploration from the bottom (less effective generators) to the top (more effective generators) of the fuzzer space. We illustrate these concepts using the example SUT in~\autoref{lst:sut-example}.

\begin{lstfloat}[t]
    \begin{lstlisting}[language=Python, escapechar=|]
def balanced_parenthesis(parens: str) -> bool:|\label{l:sut_1}|
    stack = []|\label{l:sut_2}|
    for c in parans:|\label{l:sut_3}|
        if c == "(":|\label{l:sut_4}|
            stack.insert(0, c)|\label{l:sut_5}|
        elif c == ")":|\label{l:sut_6}|
            if stack and stack[0] == "(":|\label{l:sut_7}|
                stack.pop(0)|\label{l:sut_8}|
            else:|\label{l:sut_9}|
                return False|\label{l:sut_10}|
        else:|\label{l:sut_11}|
            raise ValueError("Invalid character")|\label{l:sut_12}|
    return not stack|\label{l:sut_13}|
\end{lstlisting}
\caption{Running example}\label{lst:sut-example}
\end{lstfloat}

\begin{lstfloat}[t]
    \begin{lstlisting}[language=Python]
def fuzzer_fa(random: Random) -> str:
    random_sequence = ""
    while len(random_sequence) < MAX_LEN:
        choice = random.randint(0, CHOICE_RANGE)
        if choice == 0:
            random_sequence += "("
        elif choice == 1:
            random_sequence += ")"
        elif choice == 2:
            random_sequence += "*"
        else:
            break
    return random_sequence
\end{lstlisting}
\caption{Fuzzer $\mathrm{F}_\mathrm{A}$ covering all code of \autoref{lst:sut-example}}\label{lst:fuzzer-fa}
\end{lstfloat}

\begin{lstfloat}[t]
    \begin{lstlisting}[language=Python]
def fuzzer_fa_prime(random: Random) -> str:
    random_len = random.randint(0, MAX_LEN)
    random_sequence = ""
    for _ in range(random_len):
        choice = random.randint(0, CHOICE_RANGE)
        if choice == 0:
            random_sequence += "("
        elif choice == 1:
            random_sequence += ")"
        else:
            random_sequence += "*"
    return random_sequence
\end{lstlisting}
\caption{Fuzzer $\mathrm{F}_\mathrm{A}'$ equivalent to $\mathrm{F}_\mathrm{A}$ }\label{lst:fuzzer-fa-prime}
\end{lstfloat}

\paragraph{Cover set} For a given SUT, one important metric for characterizing fuzzers is the range of the code they cover, referred to as the \textit{cover set}. Two fuzzers with the same cover set are considered equivalent. For example, fuzzer $\mathrm{F}_\mathrm{A}$ in \autoref{lst:fuzzer-fa} and fuzzer $\mathrm{F}_\mathrm{A}'$ both generate random sequences of left and right parentheses and asterisks and cover lines~\ref{l:sut_1}--\ref{l:sut_13} of the SUT. Thus, their fuzzing effectiveness is the same. Note that the illustration uses line coverage; other forms of coverage are possible in implementation, such as edge coverage~\cite{fioraldi2020afl, wang2020not}. We use $\mathcal{C}(\cdot)$ to denote the cover set of a fuzzer.

\paragraph{The strength of fuzzers} The subset relation between cover sets defines the relative strength of fuzzers, as formalized in \autoref{def:fuzz_strength}. Intuitively, a proper superset relation of one cover set over another indicates strictly better fuzzing effectiveness. The strength of fuzzers forms a partial order, where two fuzzers can sometimes be incomparable. This occurs when the two fuzzers have distinct strengths and their cover sets cannot fully encompass each other. \autoref{lst:fuzzer-fa} and \autoref{lst:fuzzer-fb} illustrate different strengths of fuzzers. Fuzzer $\mathrm{F}_\mathrm{B}$ generates sequences of only left parentheses, covering only lines~\ref{l:sut_1}--\ref{l:sut_5} and \ref{l:sut_13}, wich is a subset of the cover set of $\mathrm{F}_\mathrm{A}$ (lines~\ref{l:sut_1}--\ref{l:sut_13}). Thus, $\mathrm{F}_\mathrm{B}$ is weaker than $\mathrm{F}_\mathrm{A}$. \looseness=-1




\begin{definition}[Strength of fuzzers]\label{def:fuzz_strength}
    Fuzzer $F_1$ is stronger than fuzzer $F_2$ if $\mathcal{C}\left(F_2\right)\subset \mathcal{C}\left(F_1\right)$, denoted as 
    \[\mathcal{S}\left(F_2\right) \sqsubset \mathcal{S}\left(F_1\right)\]
    where $\mathcal{S}(\cdot)$ is the strength of a fuzzer.
\end{definition}

\begin{figure}[t]
    \centering
    \includegraphics[scale=0.8]{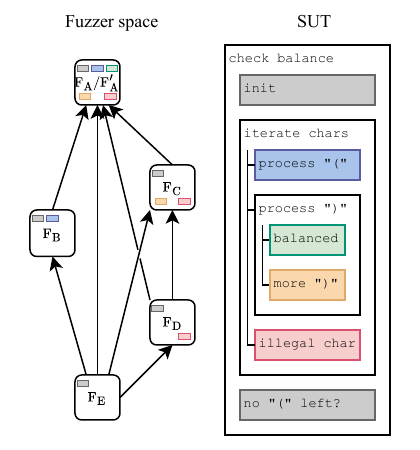}
    \caption{An illustration of \autoref{lst:sut-example}'s fuzzer space}
    \label{fig:relation-abc}
\end{figure}

\paragraph{Fuzzer space} Fuzzers with the partial order $\mathcal{S}(\cdot) \sqsubset \mathcal{S}(\cdot)$ forms a lattice, i.e., the fuzzer space. Fuzzer space enables fine-grained comparison between fuzzers. Such a comparison is impossible if using a single coverage value, which may coarsely equalize two fuzzers with distinct strengths. Intuitive correspondence exists between the lattice structure and the original fuzzers: the top element $\top$ represents the theoretically most powerful fuzzer, capable of reaching every line of code in the SUT, while the bottom element $\bot$ represents the weakest one. \autoref{fig:relation-abc} presents the fuzzer space consisting of six fuzzers, $\mathrm{F}_\mathrm{A}$~(\autoref{lst:fuzzer-fa}), $\mathrm{F}_\mathrm{A}'$~(\autoref{lst:fuzzer-fa-prime}), $\mathrm{F}_\mathrm{B}$~(\autoref{lst:fuzzer-fb}),  $\mathrm{F}_\mathrm{C}$~(\autoref{lst:fuzzer-fc}), $\mathrm{F}_\mathrm{D}$~(\autoref{lst:fuzzer-fd}), and $\mathrm{F}_\mathrm{E}$~(\autoref{lst:fuzzer-fe}). The lattice structure clearly shows the effectiveness of each fuzzer. For example, $\mathrm{F}_\mathrm{C}$ is strictly stronger than $\mathrm{F}_\mathrm{D}$, as it can cover all code covered by $\mathrm{F}_\mathrm{D}$ (the grey and red boxes) plus code that $\mathrm{F}_\mathrm{D}$ cannot cover (the yellow box). However, $\mathrm{F}_\mathrm{B}$ and $\mathrm{F}_\mathrm{C}$ cannot be thus compared, as they test different parts of the SUT. Note that if we use a single coverage value to assess the fuzzers, $\mathrm{F}_\mathrm{B}$ may be considered as replaceable by $\mathrm{F}_\mathrm{C}$, as its line coverage (which is 6) is smaller than that of $\mathrm{F}_\mathrm{C}$ (which is 11).


In real-world scenarios, generation-based fuzzers are only run for a finite time, and it is impractical and unnecessary to obtain the theoretically precise cover sets. Thus, the code covered by a fuzzer in a limited time span is used as an approximation of its cover set.

\begin{lstfloat}[t]
    \begin{lstlisting}[language=Python]
def fuzzer_fb(random: Random) -> str:
    random_sequence = ""
    while len(random_sequence) < MAX_LEN:
        choice = random.randint(0, CHOICE_RANGE)
        if choice == 0:
            random_sequence += "("
        else:
            break
    return random_sequence
\end{lstlisting}
\caption{Fuzzer $\mathrm{F}_\mathrm{B}$ weaker than $\mathrm{F}_\mathrm{A}$}\label{lst:fuzzer-fb}
\end{lstfloat}

\begin{lstfloat}[t]
    \begin{lstlisting}[language=Python]
def fuzzer_fc(random: Random) -> str:
    random_sequence = ""
    while len(random_sequence) < MAX_LEN:
        choice = random.randint(0, CHOICE_RANGE)
        if choice == 0:
            random_sequence += ")"
        elif choice == 1:
            random_sequence += "*"
        else:
            break
    return random_sequence
\end{lstlisting}
\caption{Fuzzer $\mathrm{F}_\mathrm{C}$ also weaker than $\mathrm{F}_\mathrm{A}$}\label{lst:fuzzer-fc}
\end{lstfloat}

\begin{lstfloat}[t]
    \begin{lstlisting}[language=Python]
def fuzzer_fd(random: Random) -> str:
    random_sequence = ""
    while len(random_sequence) < MAX_LEN:
        choice = random.randint(0, CHOICE_RANGE)
        elif choice == 0:
            random_sequence += "*"
        else:
            break
    return random_sequence
\end{lstlisting}
\caption{Fuzzer $\mathrm{F}_\mathrm{D}$ weaker than $\mathrm{F}_\mathrm{C}$}\label{lst:fuzzer-fd}
\end{lstfloat}

\begin{lstfloat}[t]
    \begin{lstlisting}[language=Python]
def fuzzer_fe(random: Random) -> str:
    return ''
\end{lstlisting}
\caption{Fuzzer $\mathrm{F}_\mathrm{E}$ generating only empty sequences}\label{lst:fuzzer-fe}
\end{lstfloat}

\section{LLM-driven Synthesis Over Fuzzer Space}
\label{sec:design}
\begin{figure}[t]
    \centering
    \includegraphics[width=1\linewidth]{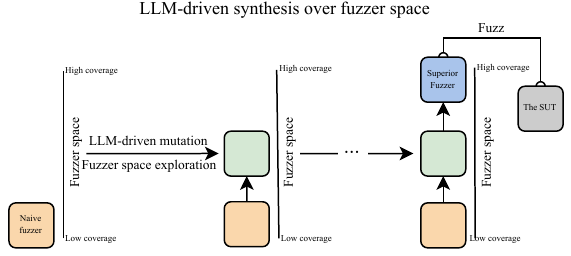}
    \caption{The workflow of \caps{ELFuzz}}
    \label{fig:elfuzz_workflow}
\end{figure}

At a high level, \sysname leverages an LLM to motivate an evolution loop to improve the seed fuzzer iteratively, and the fuzzer space steers the evolution loop towards superior fuzzers. Compared to evolutionary algorithms traditionally relying on random tree mutations~\cite{koza1994genetic}, the domain knowledge embedded in the LLM enables more reasonable and human-like modifications on the code of the candidate fuzzers~\cite{lehman2022evolution}. Then, the mutants are evaluated and placed into the fuzzer space according to their strength. Superior fuzzers closer to the top are chosen as survivors, while degenerate ones are discarded. Thus, the evolution loop ``climbs'' the fuzzer space step by step, iteratively advancing toward the optimal solution. \autoref{fig:elfuzz_workflow} illustrates this overall process.

\begin{figure}[tb]
    \centering
    \includegraphics[scale=0.9]{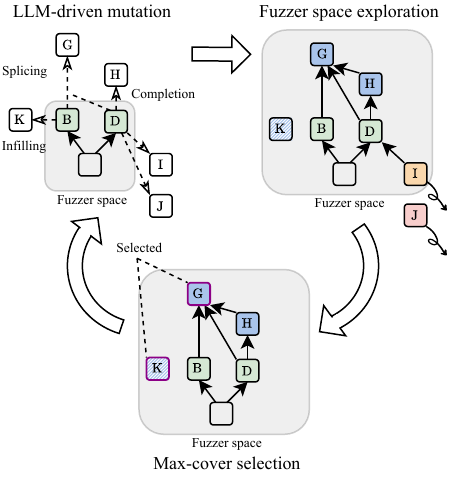}
    \caption{An evolution iteration of \caps{ELFuzz}}
    \label{fig:lefs-overview}
\end{figure}

Specifically, every iteration of the evolution loop comprises three steps as shown by \autoref{fig:lefs-overview}, each detailed by a subsection: \looseness=-1
\begin{packeditemize}
    \item \textbf{LLM-driven mutation (\S\ref{sec:mutation}).} At this step, the candidate fuzzers selected in the previous iteration as seed fuzzers are mutated by an LLM to produce mutants that may or may not augment the exploration of the fuzzer space.
    \item \textbf{Fuzzer space exploration (\S\ref{sec:fs_explore}).} At this step, the mutants are added to the already explored part of the fuzzer space according to their relative strength compared to the seed fuzzers. Invalid mutants (e.g., containing type errors) and mutants weaker than the seed fuzzers are discarded.\looseness=-1
    \item \textbf{Max-cover selection (\S\ref{sec:select}).} At this step, mutants with the maximum unioned cover set are selected from the fuzzer space as seed fuzzers for the next iteration. The selection prevents the population of candidate fuzzers from growing explosively.\looseness=-1
\end{packeditemize}

\begin{figure}[tb]
    \centering
    \includegraphics[scale=0.9]{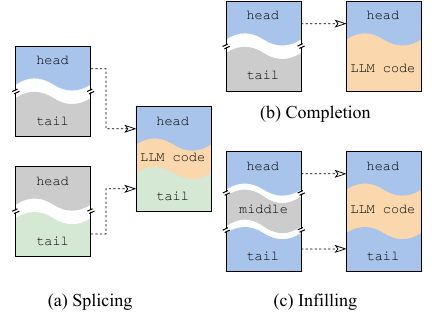}
    \caption{LLM-driven mutators} 
    \label{fig:llm_mutation}
\end{figure}


\begin{lstfloat}[t]
    \begin{lstlisting}[language=Python]
def fuzzer_fg(random: Random) -> str:
    random_sequence = ""
    while len(random_sequence) < MAX_LEN:
        choice = random.randint(0, CHOICE_RANGE)
        if choice == 0:
            random_sequence += "("
        elif choice == 1: # LLM glue code
            random_sequence += "*"
        else:
            break
    return random_sequence
\end{lstlisting}
\caption{Fuzzer $\mathrm{F}_\mathrm{G}$ stronger than the seed fuzzers}\label{lst:fuzzer-fg}
\end{lstfloat}

\begin{lstfloat}[t]
    \begin{lstlisting}[language=Python]
def fuzzer_fk(random: Random) -> str:
    random_sequence = ""
    while len(random_sequence) < MAX_LEN:
        choice = random.randint(0, CHOICE_RANGE)
        if choice == 0:
            random_sequence += "("
        else:
            break
    # Code re-completed by the LLM
    return ")" + random_sequence
\end{lstlisting}
\caption{Fuzzer $\mathrm{F}_\mathrm{K}$ discovering new code}\label{lst:fuzzer-fk}
\end{lstfloat}

\begin{lstfloat}[t]
    \begin{lstlisting}[language=Python]
def fuzzer_fh(random: Random) -> str:
    random_sequence = ""
    while len(random_sequence) < MAX_LEN:
        choice = random.randint(0, CHOICE_RANGE)
        elif choice == 0:
            # Code rewritten by the LLM
            random_sequence += "("
        else:
            break
    return random_sequence
\end{lstlisting}
\caption{Fuzzer $\mathrm{F}_\mathrm{H}$ also stronger than the seed fuzzers}\label{lst:fuzzer-fh}
\end{lstfloat}

\begin{lstfloat}[t]
    \begin{lstlisting}[language=Python]
def fuzzer_fi(random: Random) -> str:
    random_sequence = ""
    while len(random_sequence) < MAX_LEN:
        choice = random.randint(0, CHOICE_RANGE)
        elif choice == 0:
            # Code weakened by the LLM
            random_sequence += ""
        else:
            break
    return random_sequence
\end{lstlisting}
\caption{Fuzzer $\mathrm{F}_\mathrm{I}$ weaker than the seed fuzzers}\label{lst:fuzzer-fi}
\end{lstfloat}

\begin{lstfloat}[t]
    \begin{lstlisting}[language=Python]
def fuzzer_fj(random: Random) -> str:
    random_sequence = ""
    while len(random_sequence) < MAX_LEN:
        choice = random.randint(0, CHOICE_RANGE)
        elif choice == 0:
            # Type error introduced by the LLM
            random_sequence += 1
        else:
            break
    return random_sequence
\end{lstlisting}
\caption{Fuzzer $\mathrm{F}_\mathrm{J}$ containing a type error}\label{lst:fuzzer-fj}
\end{lstfloat}

\subsection{LLM-driven Mutation}\label{sec:mutation}

At the start of each iteration, the seed fuzzers are mutated by an LLM to produce mutants. The process is depicted in the leftmost part of \autoref{fig:lefs-overview}, where green boxes are the seed fuzzers and white boxes are the mutants. \sysname employs three LLM-driven mutators~\cite{lehman2022evolution} to conduct the mutation. These mutators leverage the advanced capabilities of the LLM in code generation, modification, and comprehension to apply human-like, syntactically correct, and semantically meaningful modifications. The LLM increases the likelihood of valid and diverse mutants and reduces the time spent on invalid or irrelevant ones. The details of the three mutators are provided below:
\begin{packeditemize}
    \item \textbf{Splicing.} The splicing mutator involves combining the prefix of one candidate fuzzer and the suffix of another with glue code produced via the LLM's fill-in-the-middle ability~\cite{bavarian2022efficient}. Splicing is crucial since it allows the fusion of different candidate fuzzers to join their strengths. It serves the role of crossover mutators in traditional evolutionary algorithms~\cite{langdon2013foundations, wilhelmstotterjenetics}. By gluing two fuzzers $F_1$ and $F_2$, the LLM may discover a more powerful mutant (like $F_1\sqcup F_2$) while keeping the population small. \looseness=-1

    \item \textbf{Completion.} The completion mutator involves truncating a candidate fuzzer at a random point and having the LLM complete the remaining part. Completion exploits LLM's predictive capabilities to continue an incomplete program. Note that truncation is needed because the LLM sometimes sloppily ends an incomplete program. The completion mutator enables the LLM to improve the candidate fuzzer based on previous code snippets. It mimics human programmers who implement a program line by line.

    \item \textbf{Infilling.} The infilling mutator involves removing random lines from a candidate fuzzer and utilizing the LLM's fill-in-the-middle capability to rewrite them. Infilling allows the LLM to rework parts of the code and potentially redirect the fuzzer to previously undiscovered code. Moreover, the code before and after the removed lines serves as the context for the LLM to maintain the consistency of the program~\cite{wu2024evolutionary}. Infilling mimics human programmers who revise or refactor their programs.\looseness=-1
\end{packeditemize}

We proceed to illustrate the three mutators by examples. Suppose that the previous fuzzers $\mathrm{F}_\mathrm{B}$ (\autoref{lst:fuzzer-fb}) and $\mathrm{F}_\mathrm{D}$ (\autoref{lst:fuzzer-fd}) are the seed fuzzers. The splicing mutator is applied to glue the code before line 7 of $\mathrm{F}_\mathrm{B}$ and the code after line~5 of $\mathrm{F}_\mathrm{D}$. The completion mutator is used to continue writing $\mathrm{F}_\mathrm{D}$ after removing the code after line~8. The infilling mutator is used to rewrite line~6 of $\mathrm{F}_\mathrm{B}$. The splicing mutator will cut $\mathrm{F}_\mathrm{B}$ at line~7, keeping the head and discarding the tail. The same happens to $\mathrm{F}_\mathrm{D}$ at line~5, but discarding the head and keeping the tail. Then, the head of $\mathrm{F}_\mathrm{B}$ and the tail of $\mathrm{F}_\mathrm{D}$ will be glued by the LLM to get the fuzzer $\mathrm{F}_\mathrm{G}$ in \autoref{lst:fuzzer-fg} where line~7 is the junction. The completion mutator will remove line~9 of $\mathrm{F}_\mathrm{D}$ and re-continue the code starting from this line. The fuzzer $\mathrm{F}_\mathrm{K}$ in \autoref{lst:fuzzer-fk} is the resultant mutant, while line~10 is the re-completed code by LLM. The infilling mutator will remove line~6 of $\mathrm{F}_\mathrm{B}$, resulting in a ``hole'' in the program, and the LLM will fill the hole with new code and produce the fuzzer $\mathrm{F}_\mathrm{H}$ in \autoref{lst:fuzzer-fh}, where line~7 is the code that replaces the original line~6. \looseness=-1

The LLM-driven mutators produce vast amounts of mutants, but not all of them are helpful for our synthesis. For example, a mutant can be weaker than the seed fuzzers before mutation like $\mathrm{F}_\mathrm{I}$ in \autoref{lst:fuzzer-fi} or contain type errors like $\mathrm{F}_\mathrm{J}$ in \autoref{lst:fuzzer-fj}. Fuzzer space exploration, detailed in the next subsection, will distinguish the mutants that improve the seed fuzzers and those that degenerate.

\subsection{Fuzzer Space Exploration} \label{sec:fs_explore}
\begin{algorithm}[t]
    \scriptsize
    \SetKwProg{Proc}{procedure}{ begin}{end}
    \SetFuncSty{plain}
    \SetArgSty{plain}
    \SetKwFunction{ApproxCov}{$\textsc{ApproxCov}_{SUT, T}$}
    \SetKwFunction{Explore}{$\textsc{Explore}_{SUT, T}$}
    \SetInd{0.5em}{0.5em}
    \DontPrintSemicolon 
    \Proc{\Explore{$\mathrm{fuzzerSpace}$, $\mathrm{mutants}$}} {
        \For{$m \gets \mathrm{mutants}$} {\label{l:alg_1_2}
            $\mathrm{cov}_m\gets$\ApproxCov{$m$}\;\label{l:alg_1_3}
            \For{$f\gets \mathrm{fuzzerSpace}$} {\label{l:alg_1_4}
                \If {$\mathrm{cov}_{f}\subset \mathrm{cov}_m$} {
                    $\mathrm{fuzzerSpace}.\mathrm{add\_fuzzer}(f, m)$\;
                    $\mathrm{fuzzerSpace}.\mathrm{add\_arrow}(f, m)$\;
                }
            }\label{l:alg_1_8}
        }
    }
    \Proc{\ApproxCov{$\mathrm{fuzzer}$}} {
        $\mathrm{fuzzer}.\mathrm{fuzz}(SUT, T)$\; \label{alg:line}\label{l:alg_approx_1}
        \Return $\mathrm{fuzzer}.\mathrm{coveredCode}$\;\label{l:alg_approx_2}
    }
    \caption{\bf Exploring the fuzzer space}\label{alg:explore}
\end{algorithm}

Mutants generated by LLM-driven mutation are then collected to augment the exploration of the fuzzer space, as depicted in the central part of \autoref{fig:lefs-overview}. The strength of the mutants and the seed fuzzers is compared to determine the lattice structure between them. Mutants weaker than the seed fuzzers (the orange box) and those containing errors (the red box) are discarded, while those stronger than the seed fuzzers (the blue boxes) or those covering new code (the hatched blue box) survive. This way, only mutants contributing to the ``climb'' toward the top are kept. \looseness=-1

\autoref{alg:explore} approximates the cover set of a mutant and use the mutant to augment the explored part of the fuzzer space accordingly. A detailed explanation of the algorithm is as follows:
\begin{packeditemize}
    \item Lines~\ref{l:alg_1_2} and \ref{l:alg_1_3} select a mutant fuzzer $m$ and invoke the \textsc{ApproxCov} subroutine to approximate its cover set;
    \item Lines~\ref{l:alg_1_4}--\ref{l:alg_1_8} compare the cover set of $m$ with existing fuzzers. It is added to the fuzzer space if being stronger than an existing fuzzer $f$ and the relation $\mathcal{S}(f)\sqsubset \mathcal{S}(m)$ is recorded by an arrow from $f$ to $m$;
    \item Lines~\ref{l:alg_approx_1} and \ref{l:alg_approx_2} in the \textsc{ApproxCov} subroutine approximate the cover set of a fuzzer by using it to fuzz the SUT in a limited time span. As mentioned in \S\ref{sec:overview}, obtaining the precise cover sets in real-world scenarios is impractical and unnecessary.\looseness=-1
\end{packeditemize}

Take the seed fuzzers $\mathrm{F}_\mathrm{B}$, $\mathrm{F}_\mathrm{D}$ and the mutants $\mathrm{F}_\mathrm{H}$--$\mathrm{F}_\mathrm{K}$ in the previous subsection as an example. After running for $T$ time, $\mathrm{F}_\mathrm{J}$ triggers type errors and is thus discarded. The cover sets of the other fuzzers are: 1) lines \ref{l:sut_1}--\ref{l:sut_5} and \ref{l:sut_13} for $\mathrm{F}_\mathrm{B}$ producing \codespan{"("}, \codespan{"(("} and \codespan{""}, 2) lines \ref{l:sut_1}--\ref{l:sut_3} and \ref{l:sut_11}--\ref{l:sut_13} for $\mathrm{F}_\mathrm{D}$ producing \codespan{"*"}, \codespan{"**"} and \codespan{""}, 3) lines \ref{l:sut_1}--\ref{l:sut_6} and \ref{l:sut_11}--\ref{l:sut_13} for $\mathrm{F}_\mathrm{G}$ producing \codespan{"("}, \codespan{"*"} and \codespan{""}, 4) lines~\ref{l:sut_1}--\ref{l:sut_3} and \ref{l:sut_13} for $\mathrm{F}_\mathrm{I}$ producing \codespan{""}, and 5) lines \ref{l:sut_1}--\ref{l:sut_7} and \ref{l:sut_9}--\ref{l:sut_10} for $\mathrm{F}_\mathrm{K}$ producing \codespan{"("}, \codespan{")"} and \codespan{")("}.

According to the subset relations between these cover sets, $\mathrm{F}_\mathrm{G}$ and $\mathrm{F}_\mathrm{H}$ (the blue boxes in \autoref{fig:lefs-overview}) are added to the explored part of the fuzzer space where edges from $\mathrm{F}_\mathrm{B}$ and $\mathrm{F}_\mathrm{D}$ indicate that the two new fuzzers ($\mathrm{F}_\mathrm{G}$ and $\mathrm{F}_\mathrm{H}$) are stronger than the seed fuzzers ($\mathrm{F}_\mathrm{B}$ and $\mathrm{F}_\mathrm{D}$). $\mathrm{F}_\mathrm{K}$ (the hatched blue box) is also added, as it has distinct strengths, but no edge connects to it since its cover set cannot fully encompass the cover set of any other fuzzer, and vice versa. $\mathrm{F}_\mathrm{I}$ is a fuzzer even weaker than the seed fuzzers and cannot contribute to our target of synthesizing stronger fuzzers. It is thus discarded.

\subsection{Max-cover Selection}\label{sec:select}

\begin{algorithm}[t]
    \scriptsize
    \SetKwFunction{ApproxMax}{$\textsc{ApproxMax}_{M,N,T}$}
    \SetKwFunction{GreedyMax}{$\textsc{GreedyMax}_{M,N}$}
    \SetKwFunction{BestOf}{$\textsc{BestOf}$}
    \SetKwFunction{RandPick}{$\textsc{RandPick}_{N}$}
    \SetKwProg{Proc}{procedure}{ begin}{end}
    \SetFuncSty{plain}
    \SetArgSty{plain}
    \SetInd{0.5em}{0.5em}
    \DontPrintSemicolon 
    \Proc{\ApproxMax{$\mathrm{covSets}$}} {
        $\mathrm{attempts} \gets \emptyset$\;\label{l:alg_2_rep_1}
        \For{$i\gets 1,...,T$} {
            $\mathrm{attempt}_i \gets$\GreedyMax{$\mathrm{covSets}$}\;
            $\mathrm{attempts}\gets\mathrm{attempts}\cup\{\mathrm{attempt}_i\}$\;
        }\label{l:alg_2_rep_2}
        \Return \BestOf{$\mathrm{attempts}$}\;\label{l:alg_2_best}
    }
    \Proc{\GreedyMax{$\mathrm{covSets}$}} {\label{l:alg_2_greedy_1}
        $\mathrm{attempt}\gets$\RandPick{$\mathrm{covSets}$}\;
        \While{$\mathrm{attempt}.\mathrm{changed}$} {
            \For{$\mathrm{s}\in\mathrm{attempt}$}{
                \For{$\mathrm{s}'\in\mathrm{covSets}$} {
                    \If{$|\bigcup \mathrm{attempt}[\mathrm{s}'/\mathrm{s}]|>|\bigcup \mathrm{attempt}|$} {
                        $\mathrm{attempt}\gets\mathrm{attempt}[\mathrm{s}'/\mathrm{s}]$\;
                    }
                }
            }
        }
        \KwRet $\mathrm{attempt}$\;
    }\label{l:alg_2_greedy_2}
    \caption{\bf Selecting the max-cover fuzzers}\label{alg:max-cover-set}
\end{algorithm}

As more and more parts of the fuzzer space are discovered, the seed fuzzers to be mutated will increase exponentially if not filtered. Therefore, we select a fixed-size group of elites from the mutants retained after fuzzer space exploration as the seed fuzzers for the next iteration to avoid such an explosion.

The strategy is to maximize the union of the cover sets of the elites, as illustrated by the rightmost part of \autoref{fig:lefs-overview}. Specifically, when selecting $N$ elites to advance to the next iteration from a pool $V$ of $M>N$ mutants, we choose a subset $E\subset V$ according to the following formula:
\[
E = \argmax_{S\in \mathcal{P}_{N}(V)} \left|\bigcup \left\{ \mathcal{C}\left( F \right) \middle| F\in S  \right\}\right|
\]
where $\mathcal{P}_{N}(V)$ denotes all $N$-element subsets of $V$.

\markch{The optimization problem described above is the well-known set cover problem~\cite{2024set}}, which is $\mathcal{NP}$-complete. Instead of computing an exact solution, we use an approximation algorithm, detailed in~\autoref{alg:max-cover-set}, to find a near-optimal solution. The algorithm repeats greedy optimization from random start points for $T$ times (lines~\ref{l:alg_2_rep_1}--\ref{l:alg_2_rep_2}) and chooses the best solution (line~\ref{l:alg_2_best}). The greedy optimization (lines~\ref{l:alg_2_greedy_1}--\ref{l:alg_2_greedy_2}) keeps substituting elements in the candidate solution with one that can increase the unioned cover set until no such substitution exists. The greedy optimization could be trapped in local optima, but repeating it multiple times and choosing the best mitigates the probability of being trapped in local optima. The time complexity of this approximation algorithm is $O\left(TM^2\right)$.

\section{Implementation}
\label{sec:implementation}
We have implemented \sysname in Python and made the code publicly available online. The core logic for the evolutionary algorithm and fuzzer space exploration consists of 3,602 lines of code, which is lightweight compared to traditional generation-based fuzzers like Csmith~\cite{yang2011finding} (80k~LoC), the state-of-the-art grammar synthesizer Mimid~\cite{gopinath2020mining} (13k~LoC), and the semantic constraints synthesizer ISLearn~\cite{steinhofel2022input} (10k~LoC). Details of our implementation are described below.\looseness=-1

\paragraph{Selection of the LLM} \sysname is LLM-agnostic. The LLM-driven evolution loop keeps the same, whatever LLM is used to conduct the mutation. Due to resource limits, our implementation uses a local CodeLlama model with 13 billion parameters. It is a rather small model compared to those well-known ``big models'' such as GPT-4. However, even with such a small model, \sysname has shown significant advantages over existing techniques. We attribute this to the evolution loop that decomposes the fuzzer synthesis task into small steps, which is tractable even for a small model. Bigger models can be seamlessly plugged into our implementation for better performance.

\begin{figure}[t]
    \centering
    \includegraphics[scale=0.8]{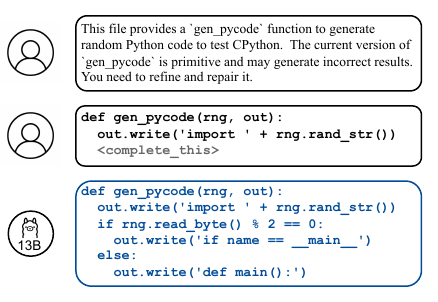}
    \caption{Prompts for the LLM-driven mutation}
    \label{fig:prompt}
\end{figure}

\paragraph{Prompt engineering} At each evolution iteration, \sysname feeds a candidate fuzzer with a prompt that states the purpose of the code (i.e., generating inputs to test a SUT) and the task for the LLM (i.e., mutating the code) prepended to the fuzzer code as comments with a simple hint for the format (e.g., XML documents always start with \texttt{<?xml ...?>}). Since the task for the LLM (viz., completing, replacing, or splicing some code, as explained in \S\ref{sec:mutation}) in each iteration is quite simple, the prompt consists of only several sentences. This minimizes the manual labor needed for prompt engineering. \autoref{fig:prompt} presents a candidate fuzzer and the prompt (simplified for illustration).

\begin{figure}[t]
    \centering
    \includegraphics[scale=0.8]{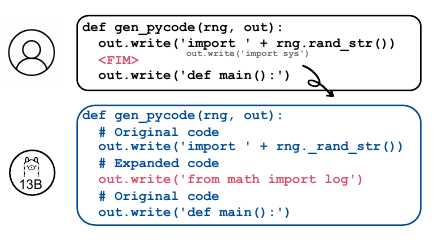}
    \caption{Fill-in-the-middle query}
    \label{fig:infill}
\end{figure}

\paragraph{Implementation of the LLM-driven mutators} The completion mutator is implemented by a text completion query to the LLM, where the LLM simply continues the queried text (i.e., the code to complete) until a terminal token. The infilling and splicing mutators are implemented via fill-in-the-middle queries~\cite{bavarian2022efficient} supported by the most recently updated LLMs. Fill-in-the-middle queries insert a special \codespanalt{FIM} token into the queried text, and the LLM will return possible expansions that fit the token's location. \autoref{fig:infill} shows how the infilling mutator is implemented via fill-in-the-middle queries. Implementation of the splicing mutator is similar: the \codespanalt{FIM} token is inserted between the head of the first candidate fuzzer and the tail of the second candidate fuzzer and expanded to glue code after the fill-in-the-middle query.

\paragraph{Seed fuzzers} \sysname improves candidate fuzzers step by step via the evolution loop. In each iteration, the LLM makes only small modifications. However, the improvements will accumulate as the process advances and eventually become a fully functional fuzzer. Thus, the initial seed fuzzers just na\"{i}vely generate purely random texts via the provided I/O utilities. The template of the seed fuzzers is included in the Appx.~\ref{sec:appx_seed}. The seed fuzzers and the synthesized fuzzers are written in Python.


\section{Evaluation}
\label{sec:eval}
We have conducted a comprehensive evaluation, which aims to investigate the following research questions on \sysname's effectiveness and characteristics:
\begin{packeditemize}
\item \textbf{RQ1: Effectiveness in fuzzing.} How effective is \sysname in fuzzing? Specifically, how much coverage can \sysname achieve compared to other input generation techniques?

\item \textbf{RQ2: Effectiveness in bug finding.} How effective is \sysname in finding bugs? How does it perform in finding bugs compared to other input generation techniques? Also, can it find new bugs in real-world software?\looseness=-1
 
\item \textbf{RQ3: Ablation study.} How do different components of \sysname, viz., the fuzzer space guidance and the three LLM-driven mutators, contribute to its effectiveness? Which is the most important one?

\item \textbf{RQ4: Interpretability and extensibility.} Do the synthesized fuzzers catch interesting characteristics of the SUTs in an interpretable way? \markch{Can the synthesized fuzzers be extended with other techniques that enhance grammar-based fuzzing?}

\end{packeditemize}

\subsection{Benchmarks and Baselines}

\paragraph{Benchmarks} Our evaluation leverages seven widely used SUTs from FuzzBench~\cite{metzman2021fuzzbench}, OSS-Fuzz~\cite{oss-fuzz} and the open-source community as the benchmarks. These SUTs span diverse fields, require structural inputs with complex grammars and/or semantic constraints, and have codebases at real-world scales (up to 1.8 million LoC).


\paragraph{Baselines} We revisited the publications listed in \autoref{tab:related_work} to select baselines. Among them, \textsc{AUTOGRAM}~\cite{hoschele2016mining}, \textsc{GLADE}~\cite{bastani2017synthesizing}, \textsc{Mimid}~\cite{gopinath2020mining}, and \textsc{ISLearn}~\cite{steinhofel2022input} share the same niche as \sysname, focusing on input generation. \textsc{AUTOGRAM}, \textsc{GLADE}, and \textsc{Mimid} synthesize grammars, while \textsc{ISLearn} synthesizes semantic constraints. Techniques in the other publications are domain-specific and/or fuse the input generation and mutation functionalities cohesively. \markch{For example, \textsc{MetaMut}~\cite{ou2024mutators} also improves fuzzing via code written by LLMs, but the application domain is strictly restricted to C compilers. They are unsuitable to serve as baselines for \sysnamehl.} \looseness=-1

\markch{As discussed in \S\ref{sec:back}, \textsc{AUTOGRAM}~\cite{gopinath2020mining} and \textsc{Mimid}~\cite{gopinath2020mining} face scalability issues. While \textsc{AUTOGRAM} is built for Java, its successor, \textsc{Mimid}, can be applied to C projects. However, the implementation of \textsc{Mimid} is tailored to the nine benchmarks in the paper, eight of which are of less than 1,000 LoC, and the remaining one is of 9,000 LoC. Our attempts showed that adapting it to the real-world SUTs of up to 1.8 million LoC as our benchmarks requires an unaffordable engineering effort. For example, \textsc{Mimid} handles only 25 kinds of AST nodes parsed by \texttt{libclang} when instrumenting C programs, but there are 248 in total~\cite{clang.cindex}.}

\markch{We include grammars synthesized by \textsc{GLADE} as a baseline, despite the doubts about the reliability of its evaluation results~\cite{bendrissou2022synthesizing}. The comparison between \textsc{GLADE} and \sysnamehl is not entirely fair due to a key implementation difference. While the \sysnamehl fuzzers synthesize test cases without references to any existing test cases, the \textsc{GLADE} fuzzer produces new test cases by mutating existing ones under the guidance of a grammar. This means \textsc{GLADE} will take advantage of the domain knowledge contained in manually written seed test cases. Besides, the grammars synthesized by \textsc{GLADE} are stored in a special binary format and cannot be leveraged by other grammar-based fuzzers.}

\markch{Therefore, another baseline is needed to manifest the effectiveness of \sysnamehl. \textsc{Grammarinator}~\cite{hodovan2018grammarinator}, a widely used grammar-based fuzzer, with manually written grammars from the \textsc{ANTLR4} project~\cite{antlr}, serves the role. The combination has two advantages to facilitate the comparison: 1) \textsc{Grammarinator} produces test cases in the same way as \sysnamehl, i.e., solely according to the grammars without references to any existing test cases, and 2) the \textsc{ANTLR4} grammars are manually written by domain experts and used as golden standards to verify the correctness of the synthesized grammars in previous work~\cite{gopinath2020mining, bendrissou2022synthesizing}, thus should reliably cover grammatical and semantic features of the SUTs.}

Besides grammars, we also want to evaluate how existing techniques perform in synthesizing semantic constraints. Semantic constraints gain less attention than grammars, as mentioned in \S\ref{sec:back}, and \textsc{ISLa}~\cite{steinhofel2022input} is the only general-purpose grammar-based fuzzer that supports semantic constraints to our knowledge. It consumes semantic constraints mined by \textsc{ISLearn}, which was proposed in the same paper. Thus, we combine these two techniques, along with the \textsc{ANTLR4} grammars, as a baseline to evaluate their performance in synthesizing and leveraging semantic constraints. We also use \textsc{ISLa}~\cite{steinhofel2022input} as a pure grammar-based fuzzer, i.e., with only the ANTLR4 grammars and without the semantic constraints, to highlight the effectiveness of the semantic constraints. Together, we get five fuzzers to compare, listed in \autoref{tab:baselines}. \looseness=-1

The format of grammars consumed by \textsc{ISLa} is different from ANTLR4, but can be translated from the latter. We implemented several Python scripts to conduct the translation.\looseness=-1

\begin{table}[tb]
    \centering
        \scriptsize
    \begin{tabularx}{\linewidth}{l>{\raggedleft\arraybackslash}p{0.8in}>{\raggedleft\arraybackslash}p{1in}>{\raggedleft\arraybackslash}X}
        \toprule
         \textbf{SUT} & \textbf{Input format} &  \textbf{Category} & \textbf{LoC}  \\
         \midrule
         \texttt{jsoncpp} & JSON & Markup language & 12,457 \\
         \texttt{libxml2} & XML & Markup language & 267,880\\
         \texttt{re2} & RegEx & Text processing & 37,431 \\
         \texttt{SQLite} & SQL & Database & 418,823 \\
         \texttt{CPython} & Python & Interpreter & 1,791,104 \\
         \texttt{cvc5} & SMT-LIB 2 & SMT solver & 553,771 \\
         \texttt{librsvg} & SVG & Image rendering & 52,125 \\
         \bottomrule
    \end{tabularx}
    \caption{Benchmarks}
    \label{tab:benchmarks}
\end{table}

\begin{table}[tb]
    \centering
        \scriptsize
    \begin{tabularx}{\linewidth}{l>{\raggedleft\arraybackslash}X}
        \toprule
        \textbf{Fuzzer} & \textbf{Short name} \\
        \midrule
        \textsc{Grammarinator} with \textsc{ANTLR4} grammars & \textsc{Grmr} \\
        \textsc{ISLa} with \textsc{ANTLR4} grammars & \textsc{ISLa} \\
        \textsc{ISLa} with \textsc{ANTLR4} grammars and \textsc{ISLearn} constraints & \textsc{ISLearn} \\
        \textsc{GLADE} & \\
         \midrule
         \sysname & \\
         \bottomrule
    \end{tabularx}
    \caption{Fuzzers to compare}
    \label{tab:baselines}
\end{table}

\subsection{Evaluation Design}

\paragraph{Answering RQ1}
\markch{To answer RQ1, we ran each fuzzer for 10 minutes to generate test cases. The generated test cases are then minimized (by \texttt{afl-cmin}) and fed to \textsc{AFL++} as seeds for 24-hour mutation-based fuzzing. Such a setup aligns with the typical usage of generation-based fuzzers, i.e., as input generators for a mutation-based fuzzer. The mutation-based fuzzing is repeated 10 times to mitigate random disturbance. We use the edge coverage to evaluate the effectiveness of fuzzing. The 10-minute setting for input generation enables the fuzzers to cover various features of the SUTs while keeping the corpus small enough for \textsc{AFL++} to explore well. The 24-hour and 10-repetition settings for mutation-based fuzzing inherit typical setups from previous work~\cite{chen2024syzgen, ammann2024dy, schloegel2024sok}.}

\paragraph{Answering RQ2} To answer RQ2, we conducted two-fold experiments, including controlled experiments and a real-world bug-finding experiment. \markch{In the controlled experiments, we fed the previously generated seeds to \textsc{AFL++} to conduct 24-hour mutation-based fuzzing to find artificial bugs injected into \texttt{libxml2}, \texttt{SQLite}, and \texttt{CPython} by \textsc{FixReverter}~\cite{zhang2022fixreverter}. \textsc{FixReverter} ``reverses'' fixes of existing CVEs and injects the reverted fixes into the software to introduce the original bugs. Such a setup provides a quantitative comparison between the fuzzers. We include only the three benchmarks written in C among all seven benchmarks because \textsc{FixReverter} only applies to C code~\cite{zhang2022fixreverter}. In total, 2,180 bugs are injected into \texttt{libxml2}, 1,092 are injected into \texttt{CPython}, and 780 are injected into \texttt{SQLite}. Each experiment is repeated 10 times to mitigate random disturbance. When a bug is triggered, the debug code attached by \textsc{FixReverter} will print the ID of the bug. We use these IDs to distinguish and count unique bugs. For example, if a test case triggers three bugs, the debug code will print information like ``Bugs 42, 43, 44 triggered'' in the standard output, and our evaluation script will count these three bugs as triggered ones.}


In the real-world bug-finding experiment, we fed the seeds generated by \sysname to \textsc{AFL++} to fuzz \texttt{cvc5} of the newest version as we started the experiment (version 1.1.2) for 14 days with 30 parallel instances to check whether \sysname can find real-world bugs.

\paragraph{Answering RQ3} To answer RQ3, we construct four variants of \sysname, each excluding a component of the approach:
\begin{packeditemize}
    \item \textsc{ELFuzz-noFS} replaces the fuzzer space model with a simple top-$k$ elite selection strategy~\cite{lalejini2022artificial} where $k$ is set to 10. In each evolution cycle, it selects the top 10 high-coverage variants without the guidance of the fuzzer space.
    \item The other three variants respectively exclude one LLM-driven mutator, viz., \caps{ELFuzz-noSP} excluding splicing, \caps{ELFuzz-noIN} excluding infilling, and \caps{ELFuzz-noCP} excluding completion.\looseness=-1
\end{packeditemize}
We recorded the coverage trends of the candidate fuzzer during the evolution to show the impact of different components on the efficiency of the evolution loop. Then, we ran the synthesized fuzzers for 10 minutes and counted the edge coverage of generated test cases to inspect the impact of different components on the effectiveness of the synthesized fuzzers. \looseness=-1

\paragraph{Answering RQ4} \markch{To answer RQ4, we conducted two case studies. The two case studies serve as evidence of the interpretability and extensibility of \sysnamehl. Although more rigorous experiments and analyses are needed to generalize the conclusions to a broader scope in the future, these case studies aim to provide initial insights within a reasonable workload and timeframe.} \looseness=-1

The first case study is to show the interpretability of \sysname. It fed 31,200 test cases sampled from the total of 3,120,000 generated by the fuzzer synthesized by \sysname to \texttt{SQLite} and counted the number of test cases that hit each source file. Fewer hits should indicate more interesting code (e.g., special \texttt{SQLite} features in contrast to commonly seen command-line option validation) in a source file. We present example test cases that hit these interesting source files and the fuzzer code that produces them. \looseness=-1

\markch{The second case study is to show the extensibility of \sysnamehl. It comprises a proof-of-concept prototype that enhances the \sysnamehl fuzzer for \texttt{cvc5} with \textsc{Zest}~\cite{padhye2019semantic}, a technique to enlarge the coverage of existing generation-based fuzzers. We will explain the crux of \textsc{Zest} and how to implement it on fuzzers synthesized by \sysnamehl. We will present the amount of work spent on implementing such a prototype and validate the correctness of the implementation.}

\subsection{Experiment Settings and Costs}

\paragraph{The LLM} We use a local full-precision (BF-16) version of CodeLlama-13B model, configured with a temperature of 0.2, a repetition penalty of 1.15, and a limit of 8192 input tokens in total. The LLM-driven evolution is run for 50 iterations for each benchmark, and each iteration produces 200 mutants, among which 10 are selected as survivors.

\paragraph{Synthesis costs} \markch{The experiments (including the synthesis processes and the fuzzing campaigns) occupy two AMD EPYC 7251 CPUs, one NVIDIA H100 Tensor Core GPU, and 16 GiB of memory. Each fuzzing campaign is assigned a single CPU core. There is no API invocation cost since all the computation happens locally. We use the NVIDIA H100 Tensor Core GPU as it is our only GPU for general-purpose computation. CodeLlama-13B is a relatively small model. Any GPU with more than 26 GiB VRAM (like NVIDIA A40~\cite{nvidia}) should be able to run it~\cite{2023codellama}, and model quantization may enable it on GPUs with even less VRAM~\cite{polino2018model}. Similarly, the 16 GiB of memory is an upper bound, too. We set the memory limit to this fixed value to ensure that it is large enough for all experiments to avoid exceptional failure. During the actual synthesis process, this upper bound has never been reached. A smaller memory limit may still fit the requirements. The synthesis of grammars for \textsc{GLADE} and semantic constraints for \textsc{ISLearn} is conducted on the same machine. They do not leverage the GPU, though.}

\markch{The time for \sysnamehl to synthesize the fuzzers is listed in \autoref{tab:time_cost}, together with the time for \textsc{GLADE} to synthesize the grammars and \textsc{ISLearn} to mine the semantic constraints. There is no data point of \textsc{ISLearn} for the \texttt{jsoncpp} benchmark, as the JSON format is simple and can be completely specified by a context-free grammar without semantic constraints. Thus, \textsc{ISLearn} does not apply to it.}

When using GLADE, we followed the replication study~\cite{bendrissou2022synthesizing} to filter out seed test cases (required by \textsc{GLADE} as ground truths~\cite{bastani2017synthesizing}) larger than 100 bytes and keep at most 30 ones randomly sampled (if there are more in total) to make the synthesis process finish in an acceptable time. Our first five attempts on \texttt{jsoncpp} without such filtration always failed due to a stack overflow error. Inspection showed that this was caused by very deep recursions on large seed test cases.

\begin{table}[t]
    \centering
    \scriptsize
    \begin{tabularx}{\linewidth}{Xr@{\hspace{1.2em}}r@{\hspace{1.2em}}r@{\hspace{1.2em}}r@{\hspace{1.2em}}r@{\hspace{1.2em}}r@{\hspace{1.2em}}r}
        \toprule
         \textbf{Fuzzer} & \texttt{jsoncpp} & \texttt{libxml2} & \texttt{re2} & \texttt{CPython} & \texttt{SQLite} & \texttt{cvc5} & \texttt{librsvg} \\
         \midrule
\textsc{ELFuzz} & \markch{21.4} & \markch{25.6} & \markch{29.5} & \markch{55.5} & \markch{40.6} & \markch{54.9} & \markch{19.9} \\
\textsc{ISLearn} & \markch{N/A}  & \markch{3.12} & \markch{0.39} & \markch{7.09} & \markch{1.71} & \markch{0.09} & \markch{0.09} \\
\textsc{GLADE} & \markch{0.20} & \markch{2.53} & \markch{0.11} & \markch{1.46} & \markch{0.29} & \markch{5.72} & \markch{1.12} \\
    \bottomrule
    \end{tabularx}
    \caption{Time costs of synthesis (h)}
    \label{tab:time_cost}
\end{table}

\markch{Generally, \sysnamehl takes the longest time as it requires multiple evolution iterations. This is justifiable since:}
\begin{packeditemize}
    \item \markch{The synthesis process in \sysnamehl happens only once before the fuzzing process. It will not harm the efficiency of fuzzing at runtime. Besides, it can be reasonably completed within 2.5 days for every benchmark.} \looseness=-1
    \item \markch{\textsc{GLADE} synthesizes only grammars. Additionally, it takes advantage of the filtration of the seed test cases.}
    \item \markch{\textsc{ISLearn} synthesizes only semantic constraints. Moreover, the synthesized semantic constraints make no contribution (compared to \textsc{ISLa} without these semantic constraints) on five out of the six benchmarks, as shown in \S\ref{sec:rq1_result} and \S\ref{sec:rq2_result} later. It brings some coverage promotion to the \texttt{librsvg} benchmark. Still, the promotion is significantly lower than \sysnamehl.} \looseness=-1
\end{packeditemize}
In later experiments, we exclude synthesis time from the recorded fuzzing time, as synthesis is not part of the AFL++ fuzzing process, and the evaluation is primarily concerned with the influence of the methods on the fuzzing effectiveness. One must trade off the promotion of the fuzzing effectiveness against the extra time spent on fuzzer synthesis when applying \sysname in practice, though. For example, in later experiments of RQ2, \sysname spent 30 minutes to trigger 75\% of the total number of bugs it can trigger in the 24-hour fuzzing experiment on \texttt{libxml2}, as shown by \autoref{tab:time_quantile}, while the second-best method spent 78 minutes. However, the speed-up is less obvious when considering the synthesis time as well. This limitation could be a direction for future research, as we will discuss in \S\ref{sec:future_work}. \looseness=-1

\paragraph{Environment and miscellaneous settings} All the experiments are conducted in Docker containers. The Docker version is 24.0.7. The operating system is Ubuntu 24.04 LTS. During fuzzing, \textsc{AFL++} uses dictionaries shipped with the SUTs~\cite{github, 2024gnome, 2024python} or provided by the OSS-Fuzz and FuzzBench projects~\cite{2024google, fuzzbench} for the six benchmarks other than \texttt{cvc5}. There are no pre-compiled dictionaries for \texttt{cvc5}. We collected tokens from the project's regression tests~\cite{2024cvc5} and used them as the dictionary. Edge coverage~\cite{durelli2018experimental} was used to compute the cover sets. Each cover set is approximated by feeding 1,000 inputs generated by a candidate fuzzer into the corresponding SUT.  \looseness=-1

\markch{\textsc{GLADE} requires manually written seed test cases as labeled ground truths for grammar synthesis. We extracted them from the test suite of each benchmark~\cite{github, 2024gnome, 2024cvc5, 2024python, 2024gnomea, 2025google, team2025sqlite}. It also requires a binary for each SUT as an oracle that returns 0 for test cases conforming to the correct format and non-zero values for others. We create such binaries using either the grammar parsing APIs of the SUTs} if provided~\cite{github, 2024gnome, 2024cvc5, 2024python, 2024gnomea, 2025google} \markch{or open-source parsers~\cite{2025sqlparser}.} Grammars of benchmarks except \texttt{librsvg} are fetched from the official grammar repository of the \textsc{ANTLR4} project~\cite{antlr}. While no formal grammar for SVG exists, SVG, based on XML, is essentially the XML grammar plus semantic constraints~\cite{2025cover, geroimenko2005svg}. Thus, we use the XML grammar for \texttt{librsvg}. \textsc{ISLearn} is expected to synthesize the extra semantic constraints for the SVG format. \looseness=-1

We use forked versions of \textsc{GLADE}, \textsc{ISLa}, and \textsc{ISLearn}, which fix fatal bugs and add some convenient command-line interfaces. The modification does not affect any of the main functionalities presented in the papers. The code of the forked versions is released together with the code of \sysname. For each benchmark, \textsc{ISLearn} typically proposes hundreds to thousands of candidate semantic constraints, while \textsc{ISLa} can only handle one during fuzzing. Thus, we keep only the candidate with the best fitness scores. \looseness=-1

\subsection{The Results}

\subsubsection{RQ1: Effectiveness in fuzzing}
\label{sec:rq1_result}

\begin{figure}
    \centering
    \includegraphics[scale=0.6]{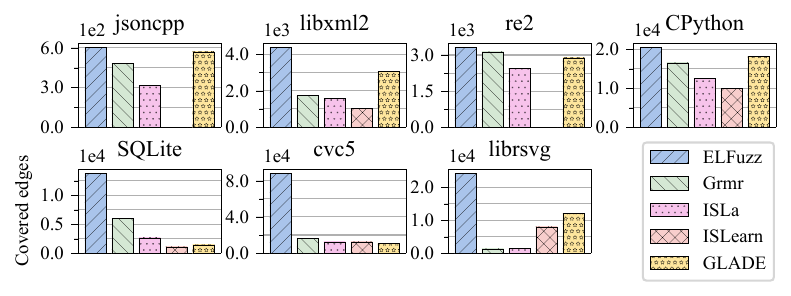}
    \caption{Coverage of the inputs generated within 10min}\label{fig:seed_cov_bar}
\end{figure}

\begin{figure}[t]
    \centering
    \includegraphics[scale=0.6]{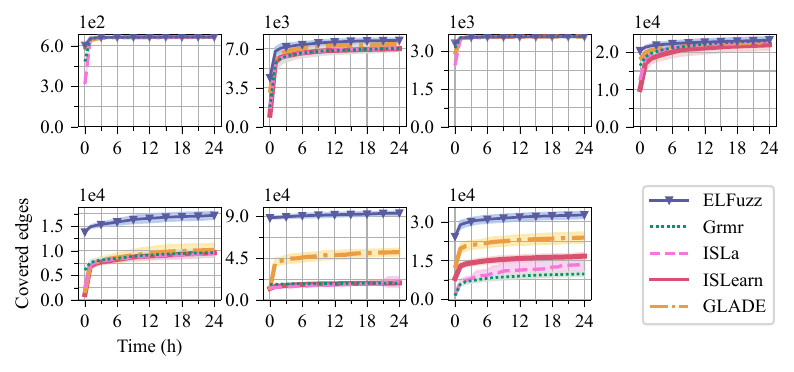}
    \caption{Coverage trends during mutation-based fuzzing}
    \label{fig:cov_trend_afl}
\end{figure}

\autoref{fig:seed_cov_bar} presents the edge coverage of inputs generated by the four fuzzers within ten minutes. For every SUT, \sysname achieves the best performance, with up to 434.8\% more covered edges over the second best (on \texttt{cvc5}). There are no results for \textsc{ISLearn} on the \texttt{re2} benchmarks because \textsc{ISLearn} failed to synthesize any semantic constraints for it (possibly because its predefined templates do not apply to the SUT). This shows the limited applicability of \textsc{ISLearn}.\looseness=-1

\markch{The coverage of seeds generated by \textsc{GLADE} is greater than that of other methods except} \sysnamehl \markch{on six out of the seven benchmarks. This is not surprising, recalling that \textsc{GLADE} gets the test cases by mutating manually written seeds, while the other methods synthesize them without references to any existing ones. Notably,} \sysnamehl \markch{significantly outperforms \textsc{GLADE} on all benchmarks, even with such unfairness.}

\autoref{fig:cov_trend_afl} presents the coverage trends of \textsc{AFL++} on seeds generated by each fuzzer to compare. \markch{The solid lines are the averaged values over the 10 repetitions, and the narrow shadows show the standard deviations.} Though the parts other than the seeds are identical in the fuzzing processes, the high coverage of seeds generated by \sysname brings significant advantages to the later fuzzing processes. On \texttt{libxml2}, \texttt{CPython}, \texttt{SQLite}, \texttt{cvc5}, and \texttt{librsvg}, \textsc{AFL++} instances that use other seeds can never catch up to the instance that uses the \sysname seeds, and the gaps stay significant (up to 436.4\% compared to the second-best results) at the end of the experiments. On \texttt{jsoncpp} and \texttt{re2}, other techniques gradually reach similar coverage as time elapses. This is because the JSON format and regular expressions both have relatively simple grammars. The grammatical tokens of them are mostly single characters (e.g., ``\{'' and ``['' in JSON, and ``?'', ``*'' in regular expressions) and do not require complex rules (e.g., matching tags in XML) to be valid. Mutation-based fuzzing can work well by itself, whatever the seeds are; therefore, the gulf brought by the seeds is gradually bridged when given sufficient time. However, seeds generated by \sysname reduce the time to reach high coverage. \looseness=-1

\markch{While \sysnamehl initially brings huge advantages, and the advantages remain huge after 24-hour fuzzing on \texttt{SQLite}, \texttt{cvc5}, and \texttt{librsvg}, \autoref{fig:cov_trend_afl} shows that the fuzzing process tends to level down the advantage on \texttt{libxml2} and \texttt{CPython}. This can be explained by the fact that \sysnamehl only provides initial seeds, and the mutation process is all the same across the methods. The longer the mutation-based fuzzing runs, the greater the proportion of the latter counts in the coverage. \S\ref{sec:future_work} will discuss this further.}

\smallskip
\noindent{\bf RQ1 Answer.} {\it
The above results demonstrate that {{ELFuzz}} can generate test cases with high coverage (up to 434.8\% more compared to other techniques). These test cases, as seeds, bring significant advantages for later mutation-based fuzzing compared to other techniques. \looseness=-1
}

\subsubsection{RQ2: Effectiveness in bug finding} \label{sec:rq2_result}

\begin{figure}[t]
    \centering
    \includegraphics[scale=0.6]{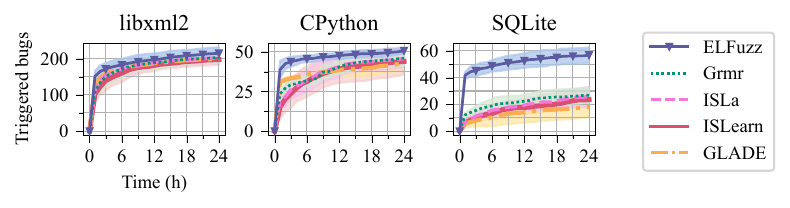}
    \caption{Number of the triggered bugs during 24-hour mutation-based fuzzing, showing that \sysname finds more bugs in a shorter time}
    \label{fig:bug}
\end{figure}

\autoref{fig:bug} presents the number of bugs triggered by \textsc{AFL++} within 24 hours using different seeds. \markch{The solid lines are the average values, and the shadows show the standard deviations.} The results demonstrate that the seeds generated by \sysname can not only find more bugs but also find the bugs faster. At the end, \sysname triggers up to 216.7\% more bugs (on \texttt{SQLite} over \textsc{GLADE}). \autoref{tab:time_quantile} presents the time (averaged over the 10 repetitions) required by the seeds generated by each fuzzer to reach 25\%, 50\%, and 75\% of the number of all bugs triggered by the seeds generated by the \sysname fuzzers as quantitative metrics to demonstrate the different speeds.\looseness=-1

\begin{table}[t]
    \centering
        \scriptsize
    \begin{tabularx}{\linewidth}{l*9{>{\raggedleft\arraybackslash}X}}
        \toprule
        \multirow{2}[2]{*}{\textbf{Fuzzer}} & \multicolumn{3}{r}{\textbf{libxml2}} & \multicolumn{3}{r}{}\textbf{CPython} & \multicolumn{3}{r}{\textbf{SQLite}} \\ 
        \cmidrule(l){2-4} \cmidrule(l){5-7} \cmidrule(l){8-10}
        & 25\% & 50\% & 75\% & 25\% & 50\% & 75\% & 25\% & 50\% & 75\% \\
         \midrule
        \textsc{ELFuzz} & \textbf{10} & \textbf{20} & \textbf{30} & \textbf{10} & \textbf{22} & \textbf{39} & \textbf{10} & \textbf{20} & \textbf{30} \\
        \textsc{Grmr} & \textbf{10} & 41 & 145 & \textbf{10} & 52 & 484 & 69 & 769 & 1080 \\
        \textsc{ISLa} & 11 & 41 & 119 & 12 & 166 & 374 & 250 & 805 & $\infty$ \\
        \textsc{ISLearn} & 11 & 57 & 173 & 78 & 264 & 434 & 332 & 1027 & $\infty$ \\
        \textsc{GLADE} & \textbf{10} & 22 & 78 & 11 & 31 & 284 & 562 & 455 & $\infty$ \\
        \midrule
        Speed-up & 1.0x & 1.1x & 2.6x & 1.0x & 1.4x & 7.3x & 6.9x & 22.8x & 36.0x \\         \bottomrule
    \end{tabularx}
    \caption{Time (min) for each method to trigger 25\%, 50\%, and 75\% of the total number of bugs triggered by \textsc{ELFuzz} and the speed-up by \textsc{ELFuzz} over the second best}
    \label{tab:time_quantile}
\end{table}

\begin{figure}
    \centering
    \includegraphics[scale=0.6]{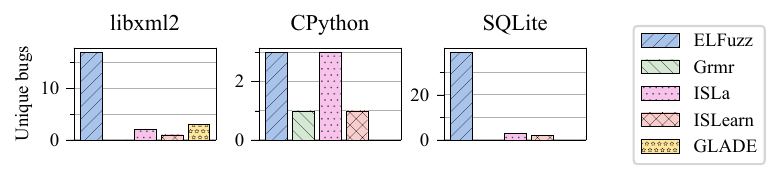}
    \caption{Unique bugs that only one fuzzer can find}
    \label{fig:unique}
\end{figure}

\markch{\autoref{fig:unique} shows the number of unique bugs (i.e., bugs that can only be triggered by the seeds generated by one specific fuzzer) they find. On \texttt{libxml2} and \texttt{SQLite}, the seeds generated by \sysnamehl find the most unique bugs (5.7--13x the second best). On \texttt{CPython}, it gets the same best number as \textsc{ISLa} (3x the third best).}

We analyzed the 16 bugs in the three benchmarks that \sysnamehl cannot trigger and found three kinds of causes. The details of the analysis are in Appx.~\ref{sec:appx_fail}. In general, six out of the 16 bugs are triggered by SUT features, e.g., \codespanalt{match} statements in Python, that are missing in the \sysnamehl fuzzers. We will discuss possible solutions to avoid such failures in \S\ref{sec:future_work}. The other ten bugs are either caused by randomness or implementation differences between \sysname and other methods. The absence of these ten bugs is not attributable to \sysname.

After fuzzing \texttt{cvc5} for 14 days by 30 AFL++ instances with the seeds generated by \sysname, five new bugs were found in the latest version of \texttt{cvc5} at the time when we started our testing. Interestingly, we noticed two of these bugs were recently fixed in the latest released version. We, therefore, have just disclosed the three unfixed bugs. Note that the five bugs include a format string injection vulnerability and two denial-of-service vulnerabilities caused by dead loops. The other two bugs involve operation on invalid pointers, which may possibly be exploited for control flow hijacking (though we have not confirmed this yet). One of the denial-of-service vulnerabilities is triggered by the complex cyclic dependency of SMT expressions generated by the synthesized fuzzer. Compared with other fuzzing papers (typically fuzzing for more than six months with more than 50 parallel AFL++ instances~\cite{ou2024mutators}), the time budget we use is short, but \sysname can still find severe and complex real-world vulnerabilities. Details of the five bugs are in Appx.~\ref{sec:appx_real}. \looseness=-1

\smallskip
\noindent{\bf RQ2 Answer.} {\it
 The above results show that {{ELFuzz}} is not only effective in covering more code but also in finding more bugs. In the controlled experiments, it finds more bugs in a shorter time than other techniques. Moreover, it can find the most unique bugs. In the real-world experiment, {{ELFuzz}} finds five new bugs, including severe and complex vulnerabilities, even with a short time budget.  \looseness=-1
}

\subsubsection{RQ3: Ablation study}
\label{sec:ablation}
\begin{figure}
    \centering
    \includegraphics[scale=0.6]{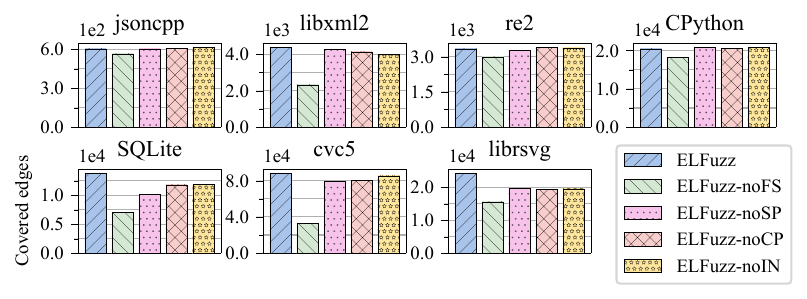}
    \caption{Coverage of the inputs generated by the variants within 10min}
    \label{fig:ablation_cov}
\end{figure}

\autoref{fig:ablation_cov} compares the coverage of test cases generated by \sysname, \caps{ELFuzz-noFS}, \caps{ELFuzz-noSP}, \caps{ELFuzz-noCP}, and \caps{ELFuzz-noIN} within 10 minutes. Among the four variants, \caps{ELFuzz-noFS} has the largest gap (6.6--62.5\%) compared to the original \sysname. This indicates that the fuzzer space model contributes the most to \sysname's performance. The three variants which exclude LLM-driven mutators (\caps{ELFuzz-noSP}, \caps{ELFuzz-noCP}, and \caps{ELFuzz-noIN}) have almost the same coverage as \sysname on the first four benchmarks, \texttt{jsoncpp}, \texttt{libxml2}, \texttt{re2}, and \texttt{CPython}. The gaps become significant on the other three benchmarks \texttt{SQLite}, \texttt{cvc5}, and \texttt{librsvg}, especially for \caps{ELFuzz-noSP} (up to 26.2\% on the three benchmarks). As stated in \S\ref{sec:mutation}, the splicing mutator is the only mutator that could combine the strengths of different candidate fuzzers. In contrast, the other two mutators can only improve a single fuzzer. Excluding the splicing mutator hinders the evolution process from joining fit features from multiple candidate fuzzers. However, none of the LLM-driven mutators are as important as the fuzzer space model. \looseness=-1

\begin{figure}
    \centering
    \includegraphics[scale=0.6]{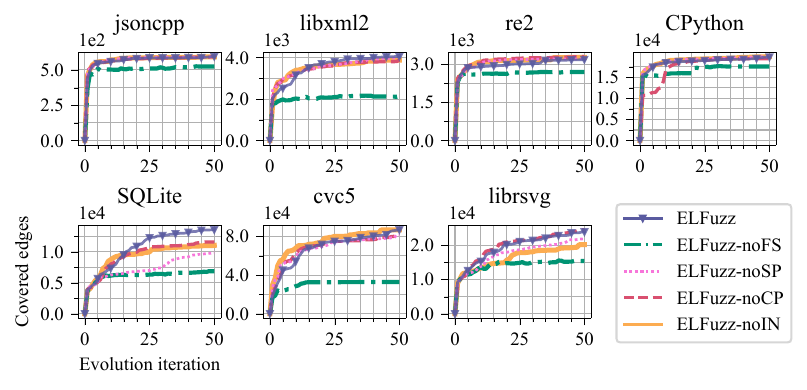}
    \caption{Edge coverage of survivors during the synthesis}
    \label{fig:ablation_trends}
\end{figure}

\autoref{fig:ablation_trends} shows the edge coverage of candidate fuzzers selected at each iteration during the LLM-driven synthesis processes, echoing the above observation that \caps{ELFuzz-noFS}'s performance downgrades the most. As the trends show, without the guidance of a fuzzer space, \caps{ELFuzz-noFS} quickly reaches the ceiling, causing significant gaps between \caps{ELFuzz-noFS} and other variants at the end of the synthesis processes. Besides, \caps{ELFuzz-noFS}'s coverage trends are not guaranteed to increase monotonically (while other variants are), as it cannot ensure that the candidate fuzzers get improved at every iteration due to the lack of guidance of the fuzzer space.\looseness=-1

\smallskip
\noindent{\bf RQ3 Answer.} {\it
Fuzzer space contributes the most to the performance of {{ELFuzz}}. While excluding an LLM-driven mutator causes some performance downgrading (up to 26.2\%), none of them have impacts as significant as that caused by excluding the fuzzer space model (up to 62.5\% 
}

\subsubsection{RQ4: Interpretability and extensibility}

In the case study on interpretability, we analyzed 31,200 test cases generated by the \sysname fuzzers for SQLite and counted the number of test cases that hit each source file. Appx.~\ref{sec:fuzzer_analysis} presents the details of the analyses. The results show a significant gap between the top-4 least-hit files and the other files: the hit numbers of the top-4 least-hit files are all less than 1,100, while those of the other files are larger than 14,000. This gap indicates interesting code that handles special SQL language features rather than trivial functionalities like command-line argument parsing in the four source files. These four files with their corresponding language features are: 1) \texttt{ctime.c} handling the \codespanalt{compile\_option} pragma, 2) \texttt{vacuum.c} handling the \codespanalt{VACUUM} command, 3) \texttt{rowset.c} operating rows in a database, and 4) \texttt{vdbesort.c} sorting the results of a SQL query. \looseness=-1

Among these language features, the \codespanalt{compile\_option} pragma and the \codespanalt{VACUUM} command are simple but rarely used. In contrast, row operations and query sorting are commonly used features with complex grammatical and semantic constraints. The synthesized fuzzers can cover both common features and rarely used features, as well as both simple features and complex features.

\autoref{lst:py_insert_row} presents part of the fuzzer code that targets \texttt{rowset.c}. The row operations in \texttt{rowset.c} requires non-empty tables, and the code satisfies this constraint by inserting rows into the generated table before invoking any row operations. Similar code exists for the other three features. Such code shows that \sysname handles the grammatical and semantic constraints of the SUTs in an interpretable way.

In the case study on extensibility, we implemented a proof-of-concept prototype that enhances the fuzzers synthesized for \texttt{cvc5} with \textsc{Zest}~\cite{padhye2019semantic}, a technique that enlarges the coverage of existing generation-based fuzzers. The crux of \textsc{Zest} is to ``parameterize'' generation-based fuzzers to let them generate test cases according to provided byte arrays instead of arbitrary random values, and then evolve the byte sequences using a genetic algorithm. Fuzzers synthesized by \sysnamehl are naturally easy to adapt to this technique, as they are basically Python functions. The adaptation can be achieved by replacing the random number generators passed to the functions as an argument with a byte array. In total, the adaptation takes five man-days. We also validated the correctness of the adaptation by manually tracing how the first 10 test cases are generated to ensure that the byte-array evolution algorithm is followed. Note that the implementation lacks various optimizations mentioned in the paper and thus cannot be a practical fuzzer due to low fuzzing efficiency. It can serve as an illustration of the extensibility of \sysnamehl, though. Appx.~\ref{sec:appx_zest} details the implementation and validation of the prototype. \looseness=-1

\begin{lstfloat}[t]
    \begin{lstlisting}[language=Python,escapechar=|]
for value in VALUES:
    testcase.write('INSERT INTO ')
    testcase.write(TABLE_NAME)|\label{l:example_13}|
    testcase.write(' VALUES (')
    for v in value[:-1]:
        testcase.write(v)
        testcase.write(', ')
        
    testcase.write(value[-1])
    testcase.write(');\n')
    \end{lstlisting}
    \caption{Synthesized code that inserts rows to a table\vspace{-1em}}
    \label{lst:py_insert_row}
\end{lstfloat}

\smallskip
\noindent{\bf RQ4 Answer.} {\it
The results above confirm that fuzzers synthesized by {{ELFuzz}} catch diverse and correct grammatical and semantic requirements of the input format of the SUT, even those complex ones, in an interpretable way. Moreover, these fuzzers can be extended with existing techniques that enhance generation-based fuzzing.\looseness=-1
}

\section{Limitations and Future Work}
\label{sec:future_work}
\paragraph{Better seed fuzzers} The current implementation of \sysname uses the same seed fuzzer that outputs random bytes for all SUTs for simplicity. Better-designed seed fuzzers, possibly produced by an LLM~\cite{liu2024oss-fuzz-gen}, may further improve the performance of \sysname. \looseness=-1

\paragraph{Data contamination and input formats} \markch{Data contamination is a well-known problem for LLM benchmarks~\cite{magar2022data}. However, it is exactly what we need for \sysnamehl. Instead of trying to \emph{test what an LLM knows}, \sysnamehl aims to \emph{leverage what an LLM knows}. Knowledge about the SUTs in the training set is necessary for the LLM to guess the input format during the synthesis of the fuzzers. Uncommon SUTs, such as proprietary software with limited publicly available documentation, may lack such knowledge. Besides, binary input formats could also cause problems, as they are typically excluded from the training sets of LLMs. Approaches such as prompt engineering or fine-tuning could help inject the necessary information into the LLM in such situations.}\looseness=-1

\paragraph{Evolution throughout the entire fuzzing cycle} Currently, \sysnamehl acts as the input generator for initial seeds. Thus, it mainly boosts the early stage of the fuzzing cycle. However, maintaining corpus diversity is crucial during the entire fuzzing process. Extending the use of the \sysnamehl fuzzers to the entire fuzzing cycle could benefit fuzzing processes more, but it faces a main difficulty caused by the complex dynamics of fuzzing. As the fuzzer touches different parts of a SUT, the desired properties of the corpus keep changing. For example, the importance of grammatical structures may decrease after the fuzzer proceeds to the assembly optimization passes when testing a compiler. Moreover, rarely used features initially ignored may be discovered, and covering them would require adaptation of the \sysnamehl fuzzer to cover them. A possible solution is to continuously evolve the input generators throughout the entire fuzzing cycle to keep pace with such changing needs. The continuous evolution and input generation can also make the time spent on fuzzer synthesis before the fuzzing campaigns more cost-effective. Future research may investigate how to leverage runtime feedback to facilitate continuous evolution, enabling adaptive input generation and the incorporation of newly discovered features of the SUTs while maintaining reasonable resource costs (such as GPU hours). \looseness=-1

\section{Conclusion}
\label{sec:conclusion}
In this paper, we introduce the novel notion of fuzzer space, which structures fuzzers into a lattice to assess their relative effectiveness. Based on the notion of fuzzer space, we develop \sysname to perform LLM-driven evolution with the fuzzer space guidance to synthesize fuzzers for fully automated, efficient, and interpretable input generation. \markch{The test cases generated by the synthesized fuzzers can then serve as initial seeds for a mutation-based fuzzer to reach deeper into the codebase.} We have implemented and evaluated \sysname. The results show that \sysname outperforms state-of-the-art techniques in fuzzing and bug-finding experiments. In a 14-day real-world fuzzing campaign on \texttt{cvc5}, \sysname found five new bugs, including three severe and complex vulnerabilities, despite the short time budget. We also conducted ablation studies. They show that the fuzzer space model contributes the most to the effectiveness of \sysname. \markch{Additionally, two case studies demonstrate that the synthesized fuzzers capture interesting features of a SUT in an interpretable way and can be extended with other generation-based fuzzing techniques.} \looseness=-1

\FloatBarrier
\clearpage
  \newpage

\section*{Acknowledgments}
\label{sec:acknowledgments}
We would like to thank the anonymous reviewers and our shepherd for providing their feedback that improved this work. This work was supported by NSF awards 2112471 and 2419722.

\section*{Ethics Considerations}
All the experiments, including those comparing different techniques on the bug-injected benchmarks, are conducted locally. No ethical risk exists in our work. Also, we have responsibly disclosed the new vulnerabilities (and bugs) to relevant stakeholders, and at this time of writing, the corresponding patch is still under development.

\section*{Open Science}
\label{open_science}
The open-source implementation of \sysname, along with the source code of the baselines and the experiment data, is available at \url{https://doi.org/10.5281/zenodo.15833146}.

\bibliographystyle{IEEEtranS}
\small
\bibliography{references.bib}

\renewcommand{\appendixname}{Appendix}
\appendixtitleon
\begin{appendices}
\clearpage

\newpage
\makeatletter
\setlength{\@fptop}{0pt}
\makeatother

\section{The Seed Fuzzers} \label{sec:appx_seed}
\autoref{lst:seed_fuzzer1} presents the seed fuzzer we use (simplified for illustration), where \codespanalt{FORMAT} is a placeholder for the input format required by the SUT. It provides clues for the LLM about what kinds of inputs it has to generate.

\begin{lstfloat}[hb]
    \begin{lstlisting}[language=Python]
class ByteOrTextIO:
    def __init__(self, binary_io: BinaryIO): ...
    def read_chars(self, char_count: int) -> str: ...
    def read_byte(self, size: int) -> int: ...

def gen_<FORMAT>(rng: Random, output: TextIO):
    len = rng.read_byte()
    random_text = rng.read_chars(len)
    output.write(random_text)
    \end{lstlisting}
    \caption{Template of the seed fuzzers}
    \label{lst:seed_fuzzer1}
\end{lstfloat}

\section{Failure Case Analysis}
\label{sec:appx_fail}

\markch{\autoref{tab:fail_case_cause} lists the causes of the failure cases in RQ2, i.e., those injected bugs that cannot be triggered by the fuzzers synthesized by \sysnamehl. There are 16 cases in total, and one of the failure cases (Bug 1425) is purely an implementation issue. Among the other cases, six (Bugs 204, 2030, 3043, 5088, 534, and 538) happen because of missing SUT features in the synthesized fuzzers, and nine (the other bugs) are caused by randomness. \S\ref{sec:future_work} discusses how to resolve the failure cases caused by missing features.} \looseness=-1

\begin{table}[hbt]
    \centering
    \scriptsize
    \begin{tabularx}{\linewidth}{llX}
        \toprule
        \textbf{Benchmark} & \textbf{Bug IDs} &  \textbf{Cause} \\
        \midrule
        \multirow[t]{4}{*}{\texttt{libxml2}} & 204, 2030, 3043 & \markch{These bugs are in code that handles multi-layer DTD element declarations. The feature is missing in the \sysnamehl fuzzers.} \\[0.25em]
        & 1425 & \markch{This bug is triggered by UTF-16 codes. The implementation of \sysnamehl forces all test cases to be UTF-8 encoded, while the \textsc{ANTLR4} grammars permit UTF-16.} \\[0.25em]
        & 3726 & \markch{This bug is in a utility function that caches used strings. The \sysnamehl fuzzers and the \textsc{ANTLR4} grammars can both generate test cases that trigger it.} \\[0.25em]
        & 4658 & \markch{The bug happens in the code that handles DTD ID declarations. The \sysnamehl fuzzers and the \textsc{ANTLR4} grammars can both generate these definitions.} \\[0.25em]
        \hdashline[2.5pt/5pt]\noalign{\vskip 0.25em}
        \multirow[t]{3}{*}{\texttt{CPython}} & 5063, 5094, 5695 & \markch{These three bugs are in code that handles keyword arguments. The \sysnamehl fuzzers and the \textsc{ANTLR4} grammars can both generate test cases containing this feature.} \\[0.25em]
        & 5078 & \markch{The bug happens during calling the \texttt{warnings} module. Neither \sysnamehl fuzzers nor the \textsc{ANTLR4} grammars specify such usage explicitly, while the mutation process may hit it by chance.} \\
        & 5088 & \markch{The bug is related to \codespanalt{match} statements, which is missing in the \sysnamehl fuzzers.} \\[0.25em]
        \hdashline[2.5pt/5pt]\noalign{\vskip 0.25em}
        \multirow[t]{3}{*}{\texttt{SQLite}} & 469 & \markch{The bug is in the code that handles the \texttt{round} function. Neither \sysnamehl fuzzers nor the \textsc{ANTLR4} grammars specify such cases explicitly, while the mutation process may hit it by chance.} \\[0.25em]
        & 474, 475 & \markch{The bugs are in the code that handles matched glob patterns. Neither \sysnamehl fuzzers nor the \textsc{ANTLR4} grammars specify such cases explicitly, while the mutation process may hit it by chance.} \\[0.25em]
        & 534, 538 & \markch{The bugs are in the code that handles nested \codespanalt{SELECT} statements. The feature is missing in the \sysnamehl fuzzers.} \\
        \bottomrule
    \end{tabularx}
    \caption{Causes of the failure cases in RQ2}
    \label{tab:fail_case_cause}
\end{table}

\section{The Real-world Bugs in \texttt{cvc5}}
\label{sec:appx_real}

\markch{Our real-world bug-finding experiment in RQ2 has found five possibly exploitable 0-day bugs in \texttt{cvc5}. These bugs are as follows:}
\begin{packeditemize}
    \item \markch{\textbf{Bug 1: Format string injection.} This bug will be triggered when passing an invalid argument to the \codespanalt{set-logic} command. When printing the error message, \texttt{cvc5} will mistakenly format it with values on the stack, possibly causing attacks that execute arbitrary commands~\cite{format}.} \looseness=-1
    \item \markch{\textbf{Bug 2: Dead loop during parsing.} This bug traps \texttt{cvc5} into a dead loop if the left parenthesis of a \codespanalt{set-option} command is not closed properly. The dead loop could lead to a denial-of-service attack.} \looseness=-1
    \item \markch{\textbf{Bug 3: Dead loop when rewriting expressions.} This bug traps \texttt{cvc5} into a dead loop when rewriting integer arithmetic expressions with self-dependency. The dead loop could lead to a denial-of-service attack.} \looseness=-1
    \item \markch{\textbf{Bug 4: Crash when setting multiple outputs.} This bug triggers a NULL-pointer dereference when setting multiple outputs.} \looseness=-1
    \item \markch{\textbf{Bug 5: Crash in \codespanalt{nullable} checks without arguments.} This bug triggers a NULL-pointer dereference when calling \codespanalt{nullable} checks without arguments.} \looseness=-1
\end{packeditemize}

\section{The Interpretability Case Study}
\label{sec:fuzzer_analysis}

\autoref{tab:least_hit} presents the hit numbers of the top-10 least-hit source files of SQLite. A significant gap exists between \texttt{vdbesort.c} and \texttt{upsert.c}. \looseness=-1

\autoref{lst:sql_row} and \autoref{lst:py_row} include a test case that covers row-related and sorting operations in SQLite and the code snippet that generates it. The test cases and code have been truncated and simplified for illustration. The test case for \texttt{ctime.c} is simply a one-line \codespanalt{compile\_option} pragma, and the synthesized fuzzer generates it if a random boolean value (the ``switch'' for this feature) is true.


\begin{table}[tb]
    \centering
        \scriptsize
    \begin{tabularx}{\linewidth}{l>{\raggedleft\arraybackslash}X>{\raggedleft\arraybackslash}X}
        \toprule
        \textbf{Source file} & \textbf{Hits} & \textbf{Percentage} \\
        \midrule
        \texttt{ctime.c} & 239 & 0.8\% \\
        \texttt{rowset.c} & 527 & 1.6\% \\
        \texttt{vacuum.c} & 707 & 2.3\% \\
        \texttt{vdbesort.c} & 1,081 & 3.5\% \\
        \texttt{upsert.c} & 14,908 & 47.8\% \\
        \texttt{random.c} & 16,030 & 51.4\% \\
        \texttt{delete.c} & 17,941 & 57.5\% \\
        \texttt{update.c} & 27,135 & 87.0\% \\
        \texttt{utf.c} & 27,135 & 87.0\% \\
        \texttt{wherecode.c} & 27,135 & 87.0\% \\
        \bottomrule
    \end{tabularx}
    \caption{The top-10 least-hit source files in \texttt{SQLite}}
    \vspace{-1em}
    \label{tab:least_hit}
\end{table}

\begin{lstfloat}[t]
    \begin{lstlisting}[language=SQL]
CREATE TABLE IF NOT EXISTS t1(
    c1, c2, c3 INTEGER PRIMARY KEY AUTOINCREMENT
);
INSERT INTO t1 VALUES (0, 1, 0);
INSERT INTO t1 VALUES (1, 2, 2);
SELECT * FROM t1 ORDER BY c1 DESC, c2 ASC, c2 DESC;
    \end{lstlisting}
    \caption{Test case for \texttt{rowset.c} and \texttt{vdbesort.c}}
    \label{lst:sql_row}
\end{lstfloat}
\begin{lstfloat}[t]
    \begin{lstlisting}[language=Python,escapechar=|]
testcase.write('CREATE TABLE IF NOT EXISTS ')
testcase.write(TABLE_NAME + '\n')
for col in COLUMN_NAMES[:-1]:
    testcase.write(col)
    testcase.write(', ')
testcase.write(COLUM_NAMES[-1])
testcase.write(
    'INTEGER PRIMARY KEY AUTOINCREMENT);\n'
)
... # Insert values into the table
orderby_columns = []
for _ in range(3):
    orderby_columns.append([
        COLUMN_NAMES[rand() % len(COLUMN_NAMES)],
        ['ASC', 'DESC'][rand() % 2]|\label{l:example_26}|
    ])

testcase.write('SELECT * FROM ')
testcase.write(TABLE_NAME)
testcase.write(' ORDER BY ')
for col in orderby_columns[:-1]:
    testcase.write(col[0])
    testcase.write(' ')
    testcase.write(col[1]))
    testcase.write(', ')

testcase.write(orderby_columns[-1][0]
testcase.write(' ')
testcase.write(orderby_columns[-1][1])
testcase.write(';\n')
    \end{lstlisting}
    \caption{Code that generates \autoref{lst:sql_row}}
    \label{lst:py_row}
\end{lstfloat}

\section{Extending \textsc{ELFuzz} with \textsc{Zest}}\label{sec:appx_zest}


\paragraph{Introduction to \textsc{Zest}} \markch{\textsc{Zest}~\cite{padhye2019semantic} is a generation-based fuzzing technique that boosts generation-based fuzzers via an evolutionary algorithm similar to greybox mutation-based fuzzers such as AFL++~\cite{fioraldi2020afl}. The crux of \textsc{Zest} is as follows:}

\markch{Traditional generation-based fuzzers are typically blind fuzzers~\cite{hodovan2018grammarinator, yang2011finding, manes2019art}, i.e., they generate test cases without feedback from the SUTs. While the specifications of the SUts enable the generation-based fuzzers to generate test cases that can pass the early parsing and semantic check stages, the lack of feedback prevents them from continuous improvements as the fuzzing process proceeds. On the contrary, greybox mutation-based fuzzers like AFL++~\cite{manes2019art, fioraldi2020afl} generate non-structural test cases but leverage coverage feedback from the SUTs to drive an evolution loop, where test cases that cover more code are preferred to produce offspring (i.e., new test cases mutated from it). Thus, the fuzzing process keeps improving itself as it proceeds.}

\markch{\textsc{Zest} proposes to graft the evolution loop of greybox mutation-based fuzzers to the generation-based fuzzers to enable the self-improvements of the latter. The fuzzer in \autoref{lst:seed_fuzzer1} is an example of a blind generation-based fuzzer, where the generated test cases solely depend on random choices according to a random number generator. \textsc{Zest} first ``parameterizes'' the fuzzer by replacing the random number generator with a caller-provided byte array, as shown in \autoref{lst:zest_elfuzz}. Thus, the byte array provided as an argument can specify the choices made while generating a test case. Then, the provided byte array can be evolved with the guidance of coverage feedback: byte arrays that result in more coverage are selected as survivors to be mutated to produce new test cases.}


\paragraph{Implementing \textsc{Zest} on \textsc{ELFuzz} fuzzers} \markch{Adapting \textsc{Zest} to the fuzzers synthesized by \textsc{ELFuzz} is straightforward: These fuzzers are evolved from the seed fuzzer in \autoref{lst:seed_fuzzer1}, and their entry points are all the same as in the list. They take a random number generator and an output stream as arguments, make choices according to the random number generator, and write the generated test case to the output. The adaptation is mainly ``parameterizing'' the fuzzers to replace the random number generator with a provided byte array, and implement the logic that evolves the byte array, as shown by \autoref{lst:zest_elfuzz}.} \looseness=-1

\begin{lstfloat}[t]
    \begin{lstlisting}[language=Python]
def collect_cov(testcase_file): ...
def looks_good(new_cov, previous_covs) -> bool: ...
def mutate_bytes(bytes) -> bytes: ...
def gen_<FORMAT>(bytes: Bytes, output: TextIO): ...

def fuzz():
  survivors = [rand_bytes()] * SURVIVOR_SIZE
  covs = []
  ... # Collect initial covs
  while True:
    select = survivors[0]
    ...
    mutant = mutate_bytes(bytes)
    gen_<FORMAT>(mutant, file)
    mut_cov = collect_cov(file)
    if looks_good(mut_cov, covs):
      survivors.pop(0)
      covs.pop(0)
      survivors.append(mutant)
      covs.append(mut_cov)
    \end{lstlisting}
    \caption{\textsc{Zest} implemented on \textsc{ELFuzz} fuzzers}
    \label{lst:zest_elfuzz}
\end{lstfloat}

\paragraph{Correctness validation} \markch{We have implemented the Zest adaptation on the \sysnamehl fuzzers for \texttt{cvc5}. We checked the first 10 test cases and their coverage information. We have confirmed that the implementation works as the above \textsc{Zest} crux demands, i.e., test cases are according to the byte arrays, and byte arrays that result in better coverage are selected and mutated to produce new test cases. For example, suppose the size of the survivor population is set to three, and test cases achieving edge coverage of 13,666, 4,336, and 24,958 are the current survivors. If the newly generated test case achieves coverage of 9,921, it will correctly replace the second test case as a new survivor.}

\paragraph{Limitations} \markch{Our current implementation of \textsc{Zest} on \sysnamehl fuzzers is only a proof-of-concept prototype to show the extensibility of \sysnamehl. We did not apply the various optimizations mentioned in the paper, which prevents the implementation from being used as a practical fuzzer. For example, we use \texttt{afl-showmap} to collect the coverage information for simplicity. At the same time, the original paper implements a binary instrumentation tool to insert coverage collection code into the SUTs. Relying on \texttt{afl-cov} brings huge overhead every time when collecting the coverage feedback.}
\end{appendices}


\newpage
\postdate{\end{center}\vspace*{-3em}}
\title{USENIX Security '25 Artifact Appendix \\ {\scshape ELFuzz}: Efficient Input Generation via LLM-driven Synthesis Over Fuzzer Space}
\author{}
\maketitle
\appendix

\section{Artifact Appendix}

This artifact appendix is a self-contained document that describes the roadmap for the evaluation of the artifacts of the paper ``{\scshape ELFuzz}: Efficient Input Generation via LLM-driven Synthesis Over Fuzzer Space.'' It includes the hardware, software, and configuration requirements for the evaluation, as well as the process for reproducing the results and claims presented in the paper. \looseness=-1

The artifacts have acquired the ``artifact available'' and ``artifact functional'' badges after the artifact evaluation. The full replication experiments for the ``artifact reproduced'' badge were not finished within the one-month time span of the artifact evaluation process due to the requirements of a significant amount of time and substantial computational resources. However, our testing shows that the artifacts function well to reproduce all the results in the paper with enough time and computational resources provided. Furthermore, we will continue to improve the artifacts for better usability and bug fixes after the artifact evaluation. \looseness=-1

\subsection{Abstract}

The artifacts contain the implementation of \sysname, the source code of the benchmarks and baselines compared with \sysname in the evaluation, the evaluation results, and the data required to reproduce the evaluation. It also includes a Docker image that preserves the environment and setups of the evaluation to facilitate one-touch replication.

\subsection{Description \& Requirements}

This section outlines the hardware and software requirements necessary to recreate the experiment environment and setups described in the paper, as well as the steps to acquire the benchmarks and baselines used in the evaluation.

\subsubsection{Security, privacy, and ethical concerns}

The artifact does not perform any operations that may compromise the security or privacy of the evaluators who execute it. \looseness=-1

\subsubsection{How to access}

The replication package, containing the code and data, is published on Zenodo. The download link of the version at the time of artifact evaluation is \url{https://doi.org/10.5281/zenodo.16741080}. Updated versions will be hosted at \url{https://doi.org/10.5281/zenodo.15833146}. The source code of \sysname and the replication package is developed in the GitHub repository at \url{https://github.com/OSUSecLab/elfuzz}.

\subsubsection{Hardware dependencies}

The experiments require a GPU with 26 GiB VRAM and CUDA support to run a local CodeLlama-13B model. The intermediate data and results produced or downloaded by the experiments will occupy approximately 100 GiB of disk space. In our evaluation, we use 30 processes to accelerate the experiments. It is recommended that the evaluators' machines be equipped with CPUs with at least this number of cores. Fewer CPU cores do not affect the results of the experiments, but a longer time to finish the experiments should be expected. The specific GPU and CPUs used in the original evaluation are one NVIDIA H100 Tensor Core GPU and two AMD EPYC 7251 CPUs. \looseness=-1

\subsubsection{Software dependencies}

The experiments do not depend on a specific operating system. A Docker installation that is compatible with version 24.0.7 is required. The host machine should also have the CUDA toolkit installed. \looseness=-1

\subsubsection{Benchmarks}

Benchmarks used in the evaluation will be automatically cloned from their official Git repositories or downloaded from the official sites during the experiments. Dockerfiles from the Fuzzbench project\footnote{\url{https://github.com/google/fuzzbench}} and the OSS-Fuzz project\footnote{\url{https://github.com/google/oss-fuzz}} are modified for replicable building environments for the six benchmarks other than \texttt{cvc5}. The Dockerfile for \texttt{cvc5} is implemented from scratch. All Dockerfiles are included in the source code tarball of the artifacts.\looseness=-1
\subsection{Setup}

This section describes how to set up and configure the environment to replicate the evaluation. It also includes instructions on the functionality validation of the replication package.

\subsubsection{Installation}

First, follow the official instructions\footnote{\url{https://docs.docker.com/engine/install/}} to install Docker. Then, set up the core pattern in the host machine as required by AFL++ later. Hereafter, ``\verb|$|'' indicates user inputs, and ``\verb|>|'' indicates program outputs.

\begin{scriptsize}
\begin{verbatim}
    $ echo core > /proc/sys/kernel/core_pattern
    > ...
\end{verbatim}
\end{scriptsize}
Now, import the Docker image and launch the container:
\begin{scriptsize}
\begin{verbatim}
    $ zstd -d elfuzz_docker_<timetag>.tar.zst
    $ docker load --input elfuzz_docker_<timetag>.tar
    $ mkdir -p /tmp/host
    $ docker run --storage-opt size=100G \
                 --cpus 30 \
                 --add-host=host.docker.internal:host-gateway \
                 -v /tmp/host:/tmp/host \
                 -v /var/run/docker.sock:/var/run/docker.sock \
                 --name elfuzz \
                 -it ghcr.io/osuseclab/elfuzz:25.07.2
\end{verbatim}
\end{scriptsize}
The commands should lead you into the container where the experiments will happen.

\subsubsection{Initilizing the environment}
\label{sec:env-setup}

After entering the Docker container, run the following commands to enable sibling containers.
\begin{scriptsize}
\begin{verbatim}
    $ sudo chown -R appuser:appuser /tmp/host/
    $ elfuzz setup
    > ...
    $ exit
\end{verbatim}
\end{scriptsize}
The two commands will exit the container. Now, restart the container for the settings to take effect.
\begin{scriptsize}
\begin{verbatim}
    $ start -ai elfuzz
\end{verbatim}
\end{scriptsize}

Then, the following command will download all the data files from Zenodo and place them in the correct locations.
\begin{scriptsize}
\begin{verbatim}
    $ elfuzz download
\end{verbatim}
\end{scriptsize}
You also need to configure your Hugging Face token:
\begin{scriptsize}
\begin{verbatim}
    $ elfuzz config --set tgi.huggingface_token <YOUR_TOKEN>
\end{verbatim}
\end{scriptsize}

\subsubsection{Functionality Validation}

After initializing the environment, you can run mini-versions of the experiments presented in the paper to validate that all of them function well. These commands adopt the same settings as the original experiments, but decrease the parameters such as time limits and the number of evolution iterations for quick validation. Each command can take 5 minutes to 2 hours to complete. \S\ref{sec:reproduce} will give detailed explanations of these commands.\looseness=-1
\begin{scriptsize}
\begin{verbatim}
    $ elfuzz synth -T fuzzer.elfuzz \
                   --use-small-model \
                   --evolution-iterations 3 \
                   jsoncpp
    $ elfuzz synth -T grammar.glade re2
    $ elfuzz synth -T semantics.islearn cvc5
    $ elfuzz produce --time 10 -T glade jsoncpp
    $ elfuzz minimize -T glade jsoncpp
    $ elfuzz run rq1.seed_cov -T glade jsoncpp
    $ elfuzz run rq1.afl --fuzzers glade \
                         --repeat 1 \
                         --time 60 \
                         jsoncpp
    $ elfuzz run rq2.afl --fuzzers glade \
                         --repeat 1 \
                         --time 60 \
                         jsoncpp
    $ elfuzz run rq2.triage
    $ elfuzz run rq2.real_world --time 60
    $ elfuzz run rq3
\end{verbatim}
\end{scriptsize}
These commands should output messages like ``ELFuzz fuzzers successfully synthesized'' without being interrupted by errors.

\subsection{Evaluation Workflow}
\label{sec:reproduce}

This section outlines the major claims presented in the evaluation part of the paper (\S\ref{sec:claim}). It describes how to conduct the experiments to fully reproduce the results presented in the paper, thereby supporting these claims (\S\ref{sec:full_scale}). However, these experiments require a significant amount of time and substantial computational resources. Thus, another option is to run the smaller-scale experiments (\S\ref{sec:full_scale}), which share the same settings as the full-scale experiments but are conducted for a shorter time or with fewer repetitions. Results produced by the small-scale experiments should also support our claims, although not 100\% replications of those in our original paper. We will explain how to conduct them and draw the same claims from the results.\looseness=-1

\subsubsection{Major claims}
\label{sec:claim}

The major claims of the paper are as follows:
\begin{packeditemize}
    \item \textbf{C1.} Fuzzers synthesized by \sysname significantly outperform the state-of-the-art generation-based fuzzers, respecting the coverage of the produced test cases and the promotion that the test cases bring to later mutation-based fuzzing processes. This is proven by the input generation experiments (E1) and the mutation-based fuzzing experiment (E2) in RQ1. \looseness=-1
    \item \textbf{C2.} Fuzzers synthesized by \sysname significantly outperform the state-of-the-art generation-based fuzzers when being used to find artificially injected bugs, and they can find previously unknown bugs in real-world SUTs. This is proved by the bug-finding experiments on bug-injected benchmarks (E3) and the real-world bug-finding experiment on \texttt{cvc5} (E4).
    \item \textbf{C3.} Fuzzer space contributes the most to the performance of \sysname among all the components. This is proved by the input generation experiments (E5) and the LLM-driven evolution processes (E6).
    \item \textbf{C4.} Fuzzers synthesized by \sysname are interpretable and extensible. This is proved by case studies on the fuzzer code (E7) and manual adaptation of \textsc{Zest}, another input-generation technique, onto the fuzzers (E8).
\end{packeditemize}

\subsubsection{Full-scale experiments}
\label{sec:full_scale}

Below are instructions on conducting full-scale experiments to replicate the results in the paper. It also lists the expected outputs and how they support our claims. Note that if you have validated the functionality of the replication package before the experiments, please reset the Docker container (remove, re-launch, and re-initialize it) to avoid data pollution caused by the results produced by the functionality validation process.
\begin{packeditemize}
    \item \textbf{E1, E5, and E6.} [1 human-hour + 10 compute-day + 50GiB disk] Inside the Docker container, use the following command to synthesize the fuzzers using \sysname or its variants (E6), or mine grammars and semantic constraints using the baselines: \looseness=-1
\begin{scriptsize}
\begin{verbatim}
    $ elfuzz synth -T <baseline> <benchmark>
\end{verbatim}
\end{scriptsize}
Please refer to the help message (\verb|--help|) on what values \verb|<baseline>| and \verb|benchmark| can take. The following commands conduct the input generation experiments (E1 and E5):
\begin{scriptsize}
\begin{verbatim}
    $ elfuzz produce -T <baseline> <benchmark>
    $ elfuzz minimize -T <baseline> <benchmark>
    $ elfuzz run rq1.seed_cov -T <baseline> <benchmark>
    $ elfuzz run rq3
\end{verbatim}
\end{scriptsize}
The coverage of the generated seed test cases will be recorded in an Excel sheet, which can be viewed via
\begin{scriptsize}
\begin{verbatim}
    $ pyexcel view /elfuzz/analysis/rq1/results/seed_cov.xlsx
\end{verbatim}
\end{scriptsize}
The coverage of seed test cases generated by \sysname is expected to be significantly higher than that generated by other methods (E1 results for C1). Similarly, you can view the coverage of seed test cases generated by \sysname variants via
\begin{scriptsize}
\begin{verbatim}
    $ pyexcel view /elfuzz/analysis/rq3/rq3_ablation.xlsx
\end{verbatim}
\end{scriptsize}
The coverage of seed test cases generated by \textsc{ELFuzz-noSP}, \textsc{ELFuzz-noCP}, and \textsc{ELFuzz-noIN} will show slight decreases, while that generated by \textsc{ELFuzz-noFS} will show significant decreases (E5 results for C3). The coverage trends of candidate fuzzers of the four variants during evolution can be viewed via
\begin{scriptsize}
\begin{verbatim}
    $ pyexcel view /elfuzz/analysis/rq3/rq3_evolve_cov.xlsx
\end{verbatim}
\end{scriptsize}
The values will display curves similar to those shown in Figure~12 of the paper if being drawn (E6 results for C3).

    \item \textbf{E2 and E3.} [1 human-hour + 20 compute-day + 5GiB disk] The following command runs the mutation-based fuzzing experiments in RQ1 (E2) and collects and analyzes the results: \looseness=-1
\begin{scriptsize}
\begin{verbatim}
    $ elfuzz run rq1.afl --fuzzers <baseline_list> \
                         --repeat 10 <benchmark_list>
\end{verbatim}
\end{scriptsize}
The following commands will run the mutation-based fuzzing campaign in RQ2 (E3) and collect and analyze the results:
\begin{scriptsize}
\begin{verbatim}
    $ elfuzz run rq2.afl --fuzzers <baseline_list> \
                         --repeat 10 <benchmark_list>
    $ elfuzz run rq2.triage
\end{verbatim}
\end{scriptsize}
You can view the averaged coverage trends during AFL++ fuzzing campaigns for RQ1 and trends of triggered bugs via
\begin{scriptsize}
\begin{verbatim}
    $ pyexcel view /elfuzz/rq1/results/rq1_sum.xlsx
    $ pyexcel view /elfuzz/rq2/results/rq2_count_bugs.xlsx
\end{verbatim}
\end{scriptsize}
The data will show curves similar to those in Figures~8 and 9, i.e., the curves of \sysname will be higher than that of others (E2 for C1 and E3 for C2).

\item \textbf{E4.} [8 human-hour + 14 compute-day] The following command runs the real-world bug-finding experiment:
\begin{scriptsize}
\begin{verbatim}
    $ elfuzz run rq2.real_world
\end{verbatim}
\end{scriptsize}
The triage and analysis of the results can only be done manually. The data tarball contains the bug-triggering test cases we found, though, in \verb|rq2/results/cvc5_bugs|. You can check whether they actually trigger the bugs of the corresponding versions of \texttt{cvc5} (E4 for C2).

\item \textbf{E7 and E8.} These two empirical case studies were done manually. The README file in the Docker image contains suggestions on how to replicate them (E7 and E8 for C4).

\item \textbf{Reproducing the figures and tables.} Use the following command to reproduce the figures and plots presented in the paper:
\begin{scriptsize}
\begin{verbatim}
    $ elfuzz plot --all
\end{verbatim}
\end{scriptsize}
Note that we cannot include proprietary fonts in the replication package. The appearance of the final rendered figures may be different from that in the original paper.
\end{packeditemize}

Please refer to the README files in the Zenodo repository, source code tarball, and Docker image tarball for more details.

\subsubsection{Small-scale experiments}
\label{sec:small_scale}
The full-scale experiments require a significant amount of time and substantial computational resources. Instead, you can choose to run some experiments at smaller scales, and the results can also support our claims. Below are instructions.
\begin{packeditemize}
    \item \textbf{E1 and E5.} [1 human-hour + 1 compute-day + 25GiB disk] Unfortunately, decreasing the number of evolution iterations will significantly affect the performance of the synthesized fuzzers, which is expected. Thus, we cannot run the small-scale versions of the synthesis processes and E6. Therefore, we will utilize the original fuzzers included in the replication package. For E1 and E5, you can use the \verb|--time 60| option for the \verb|produce| subcommand to run the seed test case generation for 1 minute instead of 10. Even this short time limit should already demonstrate the advantage of \sysname (C1) and the huge impact of excluding the fuzzer space from \sysname (C3).
    \item \textbf{E2 and E3.} [1 human-hour + 5 compute-day + 5GiB disk] Note that using seed test cases generated in a time shorter than 10 min can weaken the advantage of \sysname in AFL++ fuzzing campaigns, which is expected. Please reset the Docker container if you have run the above small-scale experiments before starting these two experiments, to ensure that you use the original test cases (generated in 10 minutes) included in the replication package. Then, you can change the value of the \verb|--repeat| option from 10 to 3 to decrease the number of repetitions. Use the \verb|--time 3600| option to run the AFL++ fuzzing campaigns for 1 hour instead of 24 hours. \sysname should be able to show advantages within the first hour of fuzzing.
\end{packeditemize}


\subsection{Version}
Based on the LaTeX template for Artifact Evaluation V20231005. Submission,
reviewing and badging methodology followed for the evaluation of this artifact
can be found at \url{https://secartifacts.github.io/usenixsec2025/}.

\end{document}